\documentclass[preprint,12pt,review]{elsarticle}

    \def\ManuscriptTitle{%
        A CFD-Based Investigation of Local Luminal Curvature as a Primary
        Determinant of Hemodynamic Environments in Cerebral Aneurysms
    }

    \journal{%
        Computers in Biology and Medicine
    }

    \usepackage{paper-style}
    \usepackage{stem-operators}

\newglossaryentry{i}{
    type=hidden,
    name={\ensuremath{i}}
}
\newglossaryentry{j}{
    type=hidden,
    name={\ensuremath{j}}
}
\newglossaryentry{k}{
    type=hidden,
    name={\ensuremath{k}}
}
\newglossaryentry{p}{
    type=hidden,
    name={\ensuremath{p}}
}
\newglossaryentry{q}{
    type=hidden,
    name={\ensuremath{q}}
}

\newglossaryentry{radial}{
    type=hidden,
    name={\ensuremath{r}},
    description={Radial coordinate of the cylindric coordinate system}
}
\newglossaryentry{azimuthal}{
    type=hidden,
    name={\ensuremath{\theta}},
    description={Circunferential coordinate of the cylindric coordinate system}
}
\newglossaryentry{z}{
    type=hidden,
    name={\ensuremath{z}},
    description={Axial coordinate of the cylindric coordinate system}
}

\newglossaryentry{xAxis}{
    type=hidden,
    name={\ensuremath{x}},
    description={%
        First coordinate of the Cartesian system (normally, $x_1$)
    }
}
\newglossaryentry{yAxis}{
    type=hidden,
    name={\ensuremath{y}},
    description={%
        Second coordinate of the Cartesian system (normally, $x_2$)
    }
}
\newglossaryentry{zAxis}{
    type=hidden,
    name={\ensuremath{z}},
    description={%
        Third coordinate of the Cartesian system (normally, $x_3$)
    }
}
\newglossaryentry{xAxisUnitVector}{
    type=hidden,
    name={\ensuremath{\vec{i}}},
    description={%
    }
}
\newglossaryentry{yAxisUnitVector}{
    type=hidden,
    name={\ensuremath{\vec{j}}},
    description={%
    }
}
\newglossaryentry{zAxisUnitVector}{
    type=hidden,
    name={\ensuremath{\vec{k}}},
    description={%
    }
}

\newglossaryentry{genericUnitVector}{
    type=hidden,
    name={\ensuremath{\vec{e}}}
}

\newglossaryentry{average}{
    type=subscript,
    name={\ensuremath{avg}},
    description={Indicates a time and surface averaged variable}
}
\newglossaryentry{maximum}{
    type=subscript,
    name={\ensuremath{max}},
    description={Indicates the maximum value of a sample}
}
\newglossaryentry{minimum}{
    type=subscript,
    name={\ensuremath{min}},
    description={Indicates the minimum value of a sample}
}
\newglossaryentry{percentil95}{
    type=subscript,
    name={\ensuremath{95}},
    description={Indicates the 95th percentile of a distribution}
}
\newglossaryentry{percentil99}{
    type=subscript,
    name={\ensuremath{99}},
    description={Indicates the 99th percentile of a distribution}
}
\newglossaryentry{solid}{
    type=superscript,
    name={\ensuremath{\mathrm{s}}},
    description={Used for solid-related variables}
}
\newglossaryentry{fluid}{
    type=superscript,
    name={\ensuremath{\mathrm{f}}},
    description={Used for fluid-related variables}
}
\newglossaryentry{fs}{
    type=superscript,
    name={\ensuremath{\mathrm{fs}}},
    description={Used for variables related to the fluid-solid interface}
}
\newglossaryentry{zeroTime}{
    type=subscript,
    name={\ensuremath{\mathrm{0}}},
    description={Indicates variables at the initial instant $t = 0$}
}
\newglossaryentry{timeIndex}{
    type=superscript,
    name={\ensuremath{\mathrm{n}}},
    description={Time index}
}
\newglossaryentry{fsiIteration}{
    type=subscript,
    name={\ensuremath{k}},
    description={%
        Index for a coupling iteration of the outer loop of the IQN-ILS
        algorithm
    }
}
\newglossaryentry{iteration}{
    type=hidden,
    name={\ensuremath{m}},
    description={%
        Index for iterations when solving a linear system  $\matrix{A}
        \cvec{x} = \cvec{b}$ with an iterative method
    }
}
\newglossaryentry{normal}{
    type=subscript,
    name={\ensuremath{n}},
    description={%
        Indicate the normal component of a vector to a surface, i.e.
        $\vec{u}_{n} = \vec{n}(\vec{n} \dprod \vec{u})$
    }
}
\newglossaryentry{tangential}{
    type=subscript,
    name={\ensuremath{t}},
    description={%
        Indicate the tangential component of a vector to a surface, i.e.
        $\vec{u}_{t} = (\tensor{I} - \vec{n}\vec{n}) \dprod \vec{u}$
    }
}
\newglossaryentry{boundary}{
    type=subscript,
    name={\ensuremath{b}},
    description={%
        Indicate a property on a boundary of a discretized domain in the
        context of the Finite Volume Method.
    }
}

\newglossaryentry{time}{
    type=latin,
    name={\ensuremath{t}},
    description={Time},
    unit={\si{\second}}
}
\newglossaryentry{positionVector}{
    type=latin,
    name={\ensuremath{\vec{r}}},
    description={Generic position vector},
    unit={\si{\meter}}
}
\newglossaryentry{temperature}{
    type=latin,
    name={\ensuremath{T}},
    description={Temperature},
    unit={\si{\kelvin}}
}
\newglossaryentry{deltat}{
    type=hidden,
    name={\ensuremath{\Delta \gls{time}}},
    description={Time-step},
    unit={\si{\second}}
}
\newglossaryentry{mass}{
    type=latin,
    name={\ensuremath{m}},
    description={Mass},
    unit={\si{\kilo\gram}}
}
\newglossaryentry{length}{
    type=latin,
    name={\ensuremath{L}},
    description={Length},
    unit={\si{\meter}}
}
\newglossaryentry{curveLength}{
    type=latin,
    name={\ensuremath{\vec{s}}},
    description={Vector length along a curve in space},
    unit={\si{\meter}}
}
\newglossaryentry{curve}{
    type=latin,
    name={\ensuremath{C}},
    description={%
        Representation of a curve in space
    },
    unit={-}
}
\newglossaryentry{perimeter}{
    type=latin,
    name={\ensuremath{P}},
    description={Perimeter of a contour},
    unit={\si{\meter}}
}
\newglossaryentry{surfaceArea}{
    type=latin,
    name={\ensuremath{A}},
    description={Area of a surface or region},
    unit={\si{\square\meter}}
}
\newglossaryentry{volume}{
    type=latin,
    name={\ensuremath{V}},
    description={Volume of a region in spatial coordinates},
    unit={\si{\cubic\meter}}
}
\newglossaryentry{surface}{
    type=latin,
    name={\ensuremath{S}},
    description={%
        Surface of volume \gls{volume}, $\partial V$, in spatial coordinates
    },
    unit={\si{\square\meter}}
}
\newglossaryentry{density}{
    type=greek,
    name={\ensuremath{\rho}},
    description={Density},
    unit={\si{\kilo\gram\per\cubic\meter}}
}
\newglossaryentry{force}{
    type=latin,
    name={\ensuremath{\vec{f}}},
    description={Generic force vector},
    unit={\si{\newton}}
}
\newglossaryentry{linearMomentum}{
    type=latin,
    name={\ensuremath{\vec{p}}},
    description={Linear momentum vector},
    unit={\si{\kilo\gram\meter\per\second}}
}
\newglossaryentry{torque}{
    type=latin,
    name={\ensuremath{\vec{T}}},
    description={Torque},
    unit={\si{\newton\meter}}
}
\newglossaryentry{angularMomentum}{
    type=latin,
    name={\ensuremath{\vec{H}}},
    description={Angular momentum vector},
    unit={\si{\kilo\gram\square\meter\per\second}}
}
\newglossaryentry{forceComp}{
    type=hidden,
    name={\ensuremath{f}},
    description={Component of the force vector},
    unit={\si{\newton}}
}
\newglossaryentry{traction}{
    type=latin,
    name={\ensuremath{\vec{t}}},
    description={Generic traction vector},
    unit={\si{\newton\per\square\meter}}
}
\newglossaryentry{tractionComp}{
    type=hidden,
    name={\ensuremath{t}},
    description={Generic traction vector},
    unit={\si{\newton\per\square\meter}}
}
\newglossaryentry{CauchyStress}{
    type=greek,
    name={\ensuremath{\tensor{\sigma}}},
    description={Cauchy stress tensor},
    unit={\si{\pascal}}
}
\newglossaryentry{CauchyStressComp}{
    type=hidden,
    name={\ensuremath{\sigma}},
    description={Cauchy stress tensor component},
    unit={\si{\pascal}}
}
\newglossaryentry{bodyForcesPerVolume}{
    type=latin,
    name={\ensuremath{\vec{f}_b}},
    description={Body forces per unit volume},
    unit={\si{\newton\per\cubic\meter}}
}
\newglossaryentry{bodyForcesPerMass}{
    type=latin,
    name={\ensuremath{\vec{b}}},
    description={Body forces per unit mass},
    unit={\si{\newton\per\kilo\gram}}
}
\newglossaryentry{bodyForcesPerMassComp}{
    type=hidden,
    name={\ensuremath{{b}}},
    description={Body forces per unit mass component},
    unit={\si{\newton\per\kilo\gram}}
}
\newglossaryentry{gravity}{
    type=latin,
    name={\ensuremath{\vec{g}}},
    description={Gravitational acceleration vector},
    unit={\si{\meter\per\square\second}}
}
\newglossaryentry{gravityComp}{
    type=hidden,
    name={\ensuremath{g}},
    description={Gravitational acceleration vector},
    unit={\si{\meter\per\square\second}}
}

\newglossaryentry{particle}{
    type=greek,
    name={\ensuremath{\xi}},
    description={Representation of a material particle},
    unit={\si{-}}
}
\newglossaryentry{EulerCoord}{
    type=latin,
    name={\ensuremath{\vec{x}}},
    description={Eulerian or spatial coordinates system (Cartesian coordinates)},
    unit={\si{\meter}}
}
\newglossaryentry{meshCoord}{
    type=greek,
    name={\ensuremath{\gvec{\chi}}},
    description={%
        Referential coordinates system for the arbitrary Lagrangian-Eulerian
        description
    },
    unit={\si{\meter}}
}
\newglossaryentry{LagrangeCoord}{
    type=greek,
    name={\ensuremath{\gvec{\xi}}},
    description={Lagrangian or material coordinates system},
    unit={\si{\meter}}
}
\newglossaryentry{meshCoordComp}{
    type=hidden,
    name={\ensuremath{\chi}},
    description={Referential, ALE or mesh coordinates system components},
    unit={\si{\meter}}
}
\newglossaryentry{EulerCoordComp}{
    type=hidden,
    name={\ensuremath{x}},
    description={Eulerian or spatial coordinates components},
    unit={\si{\meter}}
}
\newglossaryentry{LagrangeCoordComp}{
    type=hidden,
    name={\ensuremath{\xi}},
    description={Lagrangian or material coordinates components},
    unit={\si{\meter}}
}

\newglossaryentry{domain}{
    type=greek,
    name={\ensuremath{\Omega}},
    description={Spatial solution domain},
    unit={\si{\cubic\meter}}
}
\newglossaryentry{domainBoundary}{
    type=greek,
    name={\ensuremath{\Lambda}},
    description={Boundary of the domain \gls{domain}},
    unit={\si{\square\meter}}
}
\newglossaryentry{contour}{
    type=greek,
    name={\ensuremath{\Gamma}},
    description={%
        Generic three-dimensional contour; boundary of an open surface
        \gls{surface}, $\partial \gls{surface}$, in spatial coordinates
    },
    unit={\si{\meter}}
}
\newglossaryentry{nablaSpatial}{
    type=other,
    name={\ensuremath{\nabla}},
    description={Nabla operator in Eulerian coordinates},
    unit={\si{\per\meter}}
}
\newglossaryentry{nablaMaterial}{
    type=other,
    name={\ensuremath{\nabla_0}},
    description={Nabla operator in the initial material coordinates},
    unit={\si{\per\meter}}
}
\newglossaryentry{materialDomain}{
    type=greek,
    name={\ensuremath{\Omega_{\xi}}},
    description={Solution domain as seen by the material coordinates},
    unit={\si{\cubic\meter}}
}
\newglossaryentry{meshDomain}{
    type=greek,
    name={\ensuremath{\Omega_{\chi}}},
    description={Solution domain as seen by the reference coordinates},
    unit={\si{\cubic\meter}}
}
\newglossaryentry{materialVolume}{
    type=latin,
    name={\ensuremath{V_{\xi}}},
    description={Volume of an amount of particles in material coordinates},
    unit={\si{\cubic\meter}}
}
\newglossaryentry{meshVolume}{
    type=latin,
    name={\ensuremath{V_{\chi}}},
    description={Control volume in ALE coordinates},
    unit={\si{\cubic\meter}}
}
\newglossaryentry{sNormalVector}{
    type=latin,
    name={\ensuremath{\versor{n}}},
    description={%
        Unit normal vector to the surface \gls{surface} pointing outwards
    },
    unit={\si{-}}
}
\newglossaryentry{sNormalVectorComp}{
    type=hidden,
    name={\ensuremath{n}},
    unit={\si{-}}
}

\newglossaryentry{funcMaterial}{
    type=latin,
    name={\ensuremath{f_{\xi}}},
    description={Function $f$ representing any scalar variable of the material},
    unit={\si{-}}
}
\newglossaryentry{funcSpatial}{
    type=latin,
    name={\ensuremath{f}},
    description={%
        Spatial function $f$ representing any scalar variable of the material
    },
    unit={\si{-}}
}
\newglossaryentry{funcMesh}{
    type=latin,
    name={\ensuremath{f_{\chi}}},
    description={%
        ALE description of a function $f$ representing any scalar variable of
        the material
    },
    unit={\si{-}}
}
\newglossaryentry{materialSpatialMap}{
    type=greek,
    name={\ensuremath{\gvec{\Phi}}},
    description={Mapping of material coordinates onto spatial coordinates},
    unit={\si{-}}
}
\newglossaryentry{meshSpatialMap}{
    type=greek,
    name={\ensuremath{\gvec{\Psi}}},
    description={Mapping of mesh coordinates onto spatial coordinates},
    unit={\si{-}}
}
\newglossaryentry{materialMeshMap}{
    type=greek,
    name={\ensuremath{\gvec{\Xi}}},
    description={Mapping of material coordinates onto mesh coordinates},
    unit={\si{-}}
}
\newglossaryentry{identityTensor}{
    type=latin,
    name={\ensuremath{\tensor{I}}},
    description={Second order identity tensor},
    unit={\si{-}}
}
\newglossaryentry{KroneckerDeltaComp}{
    type=hidden,
    name={\ensuremath{\delta}},
    description={Kronecker's delta symbol},
    unit={\si{-}}
}
\newglossaryentry{kroneckerDelta}{
    type=hidden,
    name={\ensuremath{\gls{KroneckerDeltaComp}_{ij}}},
    description={Kronecker's delta},
    unit={\si{-}}
}
\newglossaryentry{cartesianVersor}{
    type=hidden,
    name={\ensuremath{\versor{e}}},
    description={%
        Generic cartesian versor
    }
}
\newglossaryentry{permutationTensor}{
    type=hidden,
    name={\ensuremath{\epsilon_{ijk}}},
    description={Permutation or Levi-Civita third-order tensor},
    unit={\si{-}}
}
\newglossaryentry{permutationTensorComp}{
    type=hidden,
    name={\ensuremath{\epsilon}},
    description={Permutation's tensor symbol},
    unit={\si{-}}
}
\newglossaryentry{fourthIdentityTensor}{
    type=latin,
    name={\ensuremath{\fourthTensor{I}}},
    description={Fourth-order identity tensor},
    unit={\si{-}}
}

\newglossaryentry{Jacobian}{
    type=latin,
    name={\ensuremath{J}},
    description={Jacobian (determinant of the deformation gradient)},
    unit={\si{-}}
}
\newglossaryentry{materialJacobianMatrix}{
    type=latin,
    name={\ensuremath{\mathbf{J}_{\gls{materialSpatialMap}}}},
    description={Jacobian matrix of the material-spatial mapping},
    unit={\si{-}}
}
\newglossaryentry{meshJacobianMatrix}{
    type=latin,
    name={\ensuremath{\mathbf{J}_{\gls{meshSpatialMap}}}},
    description={Jacobian matrix of the mesh-spatial mapping},
    unit={\si{-}}
}
\newglossaryentry{materialJacobian}{
    type=latin,
    name={\ensuremath{\mathrm{J}_{\gls{materialSpatialMap}}}},
    description={Jacobian of the material-spatial mapping},
    unit={\si{-}}
}
\newglossaryentry{meshJacobian}{
    type=latin,
    name={\ensuremath{\mathrm{J}_{\gls{meshSpatialMap}}}},
    description={Jacobian of the mesh-spatial mapping},
    unit={\si{-}}
}

\newglossaryentry{velocity}{
    type=latin,
    name={\ensuremath{\vec{v}}},
    description={Material velocity},
    unit={\si{\meter\per\second}}
}
\newglossaryentry{acceleration}{
    type=latin,
    name={\ensuremath{\vec{a}}},
    description={Material acceleration},
    unit={\si{\meter\per\square\second}}
}
\newglossaryentry{angularVelocity}{
    type=latin,
    name={\ensuremath{\gvec{\omega}}},
    description={Angular velocity},
    unit={\si{\radian\per\second}}
}
\newglossaryentry{velocityComp}{
    type=hidden,
    name={\ensuremath{v}},
    description={Material velocity},
    unit={\si{\meter\per\second}}
}
\newglossaryentry{angularVelocityComp}{
    type=hidden,
    name={\ensuremath{\omega}},
    unit={\si{\radian\per\second}}
}
\newglossaryentry{accelerationComp}{
    type=hidden,
    name={\ensuremath{a}},
    description={Material acceleration},
    unit={\si{\meter\per\square\second}}
}
\newglossaryentry{xVelocity}{
    type=latin,
    name={\ensuremath{v_x}},
    description={%
        Velocity component on the x direction of the spatial coordinates
    },
    unit={\si{\meter\per\second}}
}
\newglossaryentry{yVelocity}{
    type=latin,
    name={\ensuremath{v_y}},
    description={%
        Velocity component on the y direction of the spatial coordinates
    },
    unit={\si{\meter\per\second}}
}
\newglossaryentry{zVelocity}{
    type=latin,
    name={\ensuremath{v_z}},
    description={%
        Velocity component on the z direction of the spatial coordinates
    },
    unit={\si{\meter\per\second}}
}
\newglossaryentry{meshVelocity}{
    type=greek,
    name={\ensuremath{\gvec{\omega}}},
    description={Mesh velocity},
    unit={\si{\meter\per\second}}
}
\newglossaryentry{xMeshVelocity}{
    type=hidden,
    name={\ensuremath{\omega_x}},
    description={%
        Mesh velocity component on the x direction of the spatial coordinates
    },
    unit={\si{\meter\per\second}}
}
\newglossaryentry{yMeshVelocity}{
    type=hidden,
    name={\ensuremath{\omega_y}},
    description={%
        Mesh velocity component on the y direction of the spatial coordinates
    },
    unit={\si{\meter\per\second}}
}
\newglossaryentry{zMeshVelocity}{
    type=hidden,
    name={\ensuremath{\omega_z}},
    description={%
        Mesh velocity component on the z direction of the spatial coordinates
    },
    unit={\si{\meter\per\second}}
}
\newglossaryentry{meshVelocityNode}{
    type=greek,
    name={\ensuremath{\gvec{\omega}_{node}}},
    description={%
        Mesh velocity at the geometric nodes of the Finite Volume mesh
    },
    unit={\si{\meter\per\second}}
}
\newglossaryentry{materialMeshVelocity}{
    type=latin,
    name={\ensuremath{\hat{\vec{v}}}},
    description={%
        Material velocity measured from the ALE coordinates point of view
    },
    unit={\si{\meter\per\second}}
}
\newglossaryentry{convectiveVelocity}{
    type=latin,
    name={\ensuremath{\vec{c}}},
    description={Convective velocity},
    unit={\si{\meter\per\second}}
}

\newglossaryentry{numberDoF}{
    type=latin,
    name={\ensuremath{V}},
    description={Degrees of freedom of a heterogeneous thermodynamical system},
    unit={-}
}
\newglossaryentry{inlet}{
    type=subscript,
    name={\ensuremath{e}},
    description={Indicates inlet variables}
}
\newglossaryentry{outlet}{
    type=subscript,
    name={\ensuremath{s}},
    description={Indicates outlet variables}
}
\newglossaryentry{reversible}{
    type=superscript,
    name={\ensuremath{\mathrm{rev}}},
    description={Indicates a reversible process}
}
\newglossaryentry{irreversible}{
    type=superscript,
    name={\ensuremath{\mathrm{irrev}}},
    description={Indicates an irreversible process}
}
\newglossaryentry{saturated}{
    type=subscript,
    name={\ensuremath{\mathit{SAT}}},
    description={Indicates a saturated state}
}
\newglossaryentry{satLiquid}{
    type=subscript,
    name={\ensuremath{l}},
    description={Indicates the liquid-satured state}
}
\newglossaryentry{satVapor}{
    type=subscript,
    name={\ensuremath{v}},
    description={Indicates the vapor-satured state}
}
\newglossaryentry{satSolid}{
    type=subscript,
    name={\ensuremath{s}},
    description={Indicates the solid-satured state}
}
\newglossaryentry{compressedLiquid}{
    type=subscript,
    name={\ensuremath{\mathit{LC}}},
    description={Indicates compressed liquid states}
}
\newglossaryentry{critical}{
    type=subscript,
    name={\ensuremath{C}},
    description={Indicates the critical state}
}
\newglossaryentry{triple}{
    type=subscript,
    name={\ensuremath{triplo}},
    description={Indicates the triple point}
}
\newglossaryentry{reduced}{
    type=subscript,
    name={\ensuremath{r}},
    description={Indicates reduced thermodinamic properties}
}
\newglossaryentry{substance}{
    type=subscript,
    name={\ensuremath{i}},
    description={Indicates a substance}
}
\newglossaryentry{closedSystem}{
    type=subscript,
    name={\ensuremath{\textsc{s}}},
    description={Indicates a closed system quantity}
}
\newglossaryentry{controlVolume}{
    type=subscript,
    name={\ensuremath{\textsc{vc}}},
    description={Indicates a control-volume quantity}
}
\newglossaryentry{environmentState}{
    type=subscript,
    name={\ensuremath{0}},
    description={%
        Indicates a the state of the environment or surrounding of a
        system
    }
}
\newglossaryentry{hotReservoir}{
    type=subscript,
    name={\ensuremath{H}},
    description={Indicates the properties of a \enquote{hot} reservoir}
}
\newglossaryentry{coldReservoir}{
    type=subscript,
    name={\ensuremath{L}},
    description={Indicates the properties of a \enquote{cold} reservoir}
}
\newglossaryentry{cyclicMachine}{
    type=subscript,
    name={\ensuremath{M}},
    description={Indicates a cyclic thermal machine}
}
\newglossaryentry{refrigerator}{
    type=subscript,
    name={\ensuremath{R}},
    description={Indicates a refrigerator}
}
\newglossaryentry{heatPump}{
    type=subscript,
    name={\ensuremath{BC}},
    description={Indicates a heat pump}
}
\newglossaryentry{CarnotCycle}{
    type=subscript,
    name={\ensuremath{C}},
    description={Indicates a property of the cycle of Carnot}
}
\newglossaryentry{base}{
    type=hidden,
    name={\ensuremath{b}},
    description={Indicates a property at a reference arbitrary base}
}

\newglossaryentry{vaporQuality}{
    type=latin,
    name={\ensuremath{x}},
    description={%
        Vapor quality or vapor mass fraction of a saturated liquid-vapor mixture
    },
    unit={\si{\pascal}}
}
\newglossaryentry{gasConstant}{
    type=latin,
    name={\ensuremath{R}},
    description={%
        Constant of a specific gas
    },
    unit={\si{\kilo\joule\per\kilo\gram\per\kelvin}}
}
\newglossaryentry{universalGasConstant}{
    type=latin,
    name={\ensuremath{\molar{R}}},
    description={%
        Universal constant of gases
    },
    unit={\si{\kilo\joule\per\kilo\mol\per\kelvin}}
}
\newglossaryentry{compressibilityFactor}{
    type=latin,
    name={\ensuremath{Z}},
    description={Compressibiility factor},
    unit={-}
}
\newglossaryentry{PitzerAcentricFactor}{
    type=greek,
    name={\ensuremath{\omega}},
    description={Pitzer's acentric factor},
    unit={-}
}
\newglossaryentry{WuPolarFactor}{
    type=latin,
    name={\ensuremath{Y}},
    description={Wu-Stiel's polarity factor},
    unit={-}
}
\newglossaryentry{massFlowRate}{
    type=latin,
    name={\ensuremath{\dot{m}}},
    description={Mass flow rate},
    unit={\si{\kilo\gram\per\second}}
}

\newglossaryentry{speciesA}{
    type=hidden,
    name={\ensuremath{A}},
    description={Species A},
    unit={\si{-}}
}
\newglossaryentry{speciesB}{
    type=hidden,
    name={\ensuremath{B}},
    description={Species B},
    unit={\si{-}}
}
\newglossaryentry{numberMoles}{
    type=latin,
    name={\ensuremath{n}},
    description={Number of moles of a species $i$ in a system},
    unit={\si{\mol}}
}
\newglossaryentry{numberSpecies}{
    type=latin,
    name={\ensuremath{N}},
    description={Number of moles of a species $i$ in a system},
    unit={-}
}
\newglossaryentry{numberPhases}{
    type=latin,
    name={\ensuremath{p}},
    description={Number of phases in a heterogeneous mixture},
    unit={-}
}
\newglossaryentry{specificVolume}{
    type=latin,
    name={\ensuremath{v}},
    description={Specific volume -- volume per mass},
    unit={\si{\cubic\meter\per\kilogram}}
}
\newglossaryentry{modifiedReducedVolume}{
    type=hidden,
    name={\ensuremath{\gls{specificVolume}^{'}_{\gls{reduced}}}},
    description={Modified specific volume used for the reduced definition},
    unit={\si{-}}
}
\newglossaryentry{totalEnergy}{
    type=latin,
    name={\ensuremath{E}},
    description={Total energy of a system},
    unit={\si{\joule}}
}
\newglossaryentry{kineticEnergy}{
    type=latin,
    name={\ensuremath{E_c}},
    description={Kinetic energy of a system},
    unit={\si{\joule}}
}
\newglossaryentry{potentialEnergy}{
    type=latin,
    name={\ensuremath{E_p}},
    description={Potential energy of a system},
    unit={\si{\joule}}
}
\newglossaryentry{internalEnergy}{
    type=latin,
    name={\ensuremath{U}},
    description={Internal energy of a system},
    unit={\si{\joule}}
}
\newglossaryentry{intInternalEnergy}{
    type=latin,
    name={\ensuremath{u}},
    description={Intensive internal energy of a system},
    unit={\si{\joule\per\kilo\gram}}
}
\newglossaryentry{intTotalEnergy}{
    type=latin,
    name={\ensuremath{e}},
    description={Intensive total energy of a system},
    unit={\si{\joule\per\kilo\gram}}
}
\newglossaryentry{enthalpy}{
    type=latin,
    name={\ensuremath{H}},
    description={Enthalpy of a system},
    unit={\si{\joule}}
}
\newglossaryentry{intEnthalpy}{
    type=latin,
    name={\ensuremath{h}},
    description={Intensive enthalpy of a system},
    unit={\si{\joule\per\kilo\gram}}
}

\newglossaryentry{entropy}{
    type=latin,
    name={\ensuremath{S}},
    description={Entropy of a system},
    unit={\si{\joule\per\kelvin}}
}
\newglossaryentry{intEntropy}{
    type=latin,
    name={\ensuremath{s}},
    description={Intensive entropy of a system},
    unit={\si{\joule\per\kilo\gram\per\kelvin}}
}

\newglossaryentry{HelmholtzFreeEnergy}{
    type=latin,
    name={\ensuremath{A}},
    description={Helmholtz free-energy of a system},
    unit={\si{\joule}}
}
\newglossaryentry{intHelmholtzFreeEnergy}{
    type=latin,
    name={\ensuremath{a}},
    description={Intensive Helmholtz free-energy of a system},
    unit={\si{\joule\per\kilo\gram}}
}

\newglossaryentry{GibbsFreeEnergy}{
    type=latin,
    name={\ensuremath{G}},
    description={Gibbs free-energy of a system},
    unit={\si{\joule}}
}
\newglossaryentry{intGibbsFreeEnergy}{
    type=latin,
    name={\ensuremath{g}},
    description={Intensive Gibbs free-energy of a system},
    unit={\si{\joule\per\kilo\gram}}
}

\newglossaryentry{entropyCreated}{
    type=latin,
    name={\ensuremath{S_{gen}}},
    description={%
        Entropy created or generated in a process undergone by a system
    },
    unit={\si{\joule\per\kelvin}}
}
\newglossaryentry{intEntropyCreated}{
    type=latin,
    name={\ensuremath{s_{gen}}},
    description={%
        Intensive entropy created or generated in a process undergone by a
        system
    },
    unit={\si{\joule\per\kelvin}}
}
\newglossaryentry{entropyCreatedRate}{
    type=latin,
    name={\ensuremath{{\transferRate{entropy}}_{gen}}},
    description={%
        Entropy created or generated in a process undergone by a system
    },
    unit={\si{\watt\per\kelvin}}
}
\newglossaryentry{intEntropyCreatedRate}{
    type=latin,
    name={\ensuremath{\dot{s}_{gen}}},
    description={%
        Intensive entropy created or generated in a process undergone by a
        system
    },
    unit={\si{\watt\per\kelvin}}
}

\newglossaryentry{chemicalPotential}{
    type=greek,
    name={\ensuremath{\mu}},
    description={%
        Chemical potential of a species crossing the boundary of a system
    },
    unit={\si{\joule\per\kelvin}}
}
\newglossaryentry{irreversibility}{
    type=latin,
    name={\ensuremath{I}},
    description={Irreversibility of a thermodynamical process},
    unit={\si{\joule}}
}
\newglossaryentry{intIrreversibility}{
    type=latin,
    name={\ensuremath{i}},
    description={Intensive irreversibility},
    unit={\si{\joule\per\kilo\gram}}
}
\newglossaryentry{irreversibilityRate}{
    type=latin,
    name={\ensuremath{\dot{I}}},
    description={Irreversibility rate of a thermodynamical process},
    unit={\si{\watt}}
}

\newglossaryentry{thermodynamicProperty}{
    type=hidden,
    name={\ensuremath{X}},
    description={%
        Generic extensive thermodynamic property of a system
    },
    unit={-}
}
\newglossaryentry{intThermodynamicProperty}{
    type=hidden,
    name={\ensuremath{x}},
    description={%
        Generic intensive thermodynamic property of a system
    },
    unit={-}
}
\newglossaryentry{thermodynamicSubstate}{
    type=greek,
    name={\ensuremath{\upsilon}},
    description={%
        Generic thermodynamic sub-state of a system, identified as the
        intensive counterpart of the generalized displacement producing work
        transfers
    },
    unit={-}
}
\newglossaryentry{generalizedForce}{
    type=greek,
    name={\ensuremath{\tau}},
    description={Generalized force applied to a system},
    unit={-}
}
\newglossaryentry{generalizedDisplacement}{
    type=greek,
    name={\ensuremath{\Upsilon}},
    description={Generalized displacement applied to a system},
    unit={-}
}
\newglossaryentry{heatTransfer}{
    type=latin,
    name={\ensuremath{Q}},
    description={Heat transfer interaction on the boundary of a system},
    unit={\si{\joule}}
}
\newglossaryentry{thermalConductivity}{
    type=latin,
    name={\ensuremath{k}},
    description={%
        Thermal conductivity of a material
    },
    unit={\si{\watt\per\meter\per\kelvin}}
}
\newglossaryentry{thermalDiffusivity}{
    type=greek,
    name={\ensuremath{\alpha}},
    description={%
        Thermal diffusivity of a material
    },
    unit={\si{\square\meter\per\second}}
}
\newglossaryentry{specificHeatTransfer}{
    type=latin,
    name={\ensuremath{q}},
    description={%
        Heat transfer interaction on the boundary of a system per unit
        mass
    },
    unit={\si{\joule\per\kilo\gram}}
}
\newglossaryentry{heatFlux}{
    type=latin,
    name={\ensuremath{q^{"}}},
    description={%
        Unidirection heat flux
    },
    unit={\si{\watt\per\square\meter}}
}
\newglossaryentry{heatFluxVector}{
    type=latin,
    name={\ensuremath{\vec{q}^{"}}},
    description={%
        Heat flux vector
    },
    unit={\si{\watt\per\square\meter}}
}
\newglossaryentry{logMeanTemperatureDifference}{
    type=hidden,
    name={\ensuremath{\Delta T_{\text{ml}}}},
    description={%
        Log mean tempratura difference
    },
    unit={\si{\kelvin}}
}
\newglossaryentry{specificHeatGeneration}{
    type=latin,
    name={\ensuremath{\dot{q}}},
    description={%
        Rate of heat generated per unit volume in the interior of a medium
    },
    unit={\si{\watt\per\cubic\meter}}
}
\newglossaryentry{heatFlowVector}{
    type=latin,
    name={\ensuremath{\vec{q}}},
    description={%
        Heat flux or heat flow vector, defined as the heat transfer per unit
        surface area
    },
    unit={\si{\watt\per\square\meter}}
}
\newglossaryentry{molarFlux}{
    type=latin,
    name={\ensuremath{N^{"}}},
    description={%
        Molar flux of a species
    },
    unit={\si{\mol\per\second}}
}
\newglossaryentry{massFlux}{
    type=latin,
    name={\ensuremath{n^{"}}},
    description={%
        Mass flux of a species
    },
    unit={\si{\kilo\gram\per\second\per\square\meter}}
}
\newglossaryentry{molarTransferRate}{
    type=latin,
    name={\ensuremath{N}},
    description={%
        Molar transfer rate of a species
    },
    unit={\si{\mol\per\second}}
}
\newglossaryentry{molarConcentration}{
    type=greek,
    name={\ensuremath{C}},
    description={Molar concentration},
    unit={\si{\kilo\mol\per\cubic\meter}}
}
\newglossaryentry{binaryDiffusionCoefficient}{
    type=latin,
    name={\ensuremath{D}},
    description={%
        Binary diffusion coefficent
    },
    unit={\si{\square\meter\per\second}}
}

\newglossaryentry{workTransfer}{
    type=latin,
    name={\ensuremath{W}},
    description={Generic work transfer interaction on the boundary of a system},
    unit={\si{\joule}}
}
\newglossaryentry{heatTransferRate}{
    type=latin,
    name={\ensuremath{\transferRate{heatTransfer}}},
    description={Heat transfer rate},
    unit={\si{\kilo\joule\per\second}}
}
\newglossaryentry{workTransferRate}{
    type=latin,
    name={\ensuremath{\transferRate{workTransfer}}},
    description={Work transfer rate},
    unit={\si{\kilo\joule\per\second}}
}
\newglossaryentry{specifcWorkRate}{
    type=latin,
    name={\ensuremath{w}},
    description={Work transfer rate},
    unit={\si{\kilo\joule\per\kilo\gram}}
}
\newglossaryentry{internalMechanicalPower}{
    type=latin,
    name={\ensuremath{\dot{w}_{int}}},
    description={%
        Internal mechanical power in a system, caused by the stress and
        deformation
    },
    unit={\si{\watt}}
}
\newglossaryentry{efficiency}{
    type=greek,
    name={\ensuremath{\eta}},
    description={Generic efficiency of a thermal machine or other system},
    unit={-}
}
\newglossaryentry{performanceCoeff}{
    type=greek,
    name={\ensuremath{\beta}},
    description={%
        Generic coefficient of performance of a refrigerator or heat pump
    },
    unit={-}
}
\newglossaryentry{secondLawEfficiency}{
    type=greek,
    name={\ensuremath{\gls{efficiency}_{\textsc{ii}}}},
    description={Efficiency of a thermal system in terms of the Second Law},
    unit={-}
}
\newglossaryentry{numberWorkSources}{
    type=latin,
    name={\ensuremath{m}},
    description={Number of work sources to a system},
    unit={}
}

\newglossaryentry{bulkModulus}{
    type=greek,
    name={\ensuremath{\kappa}},
    description={Bulk modulus},
    unit={\si{\pascal}}
}

\newglossaryentry{satTemperature}{
    type=latin,
    name={\ensuremath{\gls{temperature}_{SAT}}},
    description={Saturation temperature of a pure substance},
    unit={\si{\kelvin}}
}
\newglossaryentry{satPressure}{
    type=latin,
    name={\ensuremath{\gls{pressure}_{SAT}}},
    description={Saturation pressure of a pure substance},
    unit={\si{\pascal}}
}

\newglossaryentry{exergy}{
    type=latin,
    name={\ensuremath{E}},
    description={Generic symbol representing exergy},
    unit={\si{\joule}}
}
\newglossaryentry{exergyRate}{
    type=latin,
    name={\ensuremath{\dot{E}}},
    description={Generic symbol representing exergy rate},
    unit={\si{\watt}}
}
\newglossaryentry{flowExergy}{
    type=latin,
    name={\ensuremath{\Psi}},
    description={Flow exergy},
    unit={\si{\joule}}
}
\newglossaryentry{nonFlowExergy}{
    type=latin,
    name={\ensuremath{\Phi}},
    description={Non-flow exergy},
    unit={\si{\joule}}
}
\newglossaryentry{intFlowExergy}{
    type=latin,
    name={\ensuremath{\psi}},
    description={Intensive flow exergy},
    unit={\si{\joule\per\kilo\gram}}
}
\newglossaryentry{intNonFlowExergy}{
    type=latin,
    name={\ensuremath{\varphi}},
    description={Intensive non-flow exergy},
    unit={\si{\joule\per\kilo\gram}}
}
\newglossaryentry{specificHeat}{
    type=latin,
    name={\ensuremath{c}},
    description={Specific heat of a pure substance},
    unit={\si{\joule\per\kilo\gram\per\kelvin}}
}
\newglossaryentry{constVolumeSpecificHeat}{
    type=latin,
    name={\ensuremath{\gls{specificHeat}_{\gls{volume}}}},
    description={Constant-volume specific heat of a pure substance},
    unit={\si{\joule\per\kilo\gram\per\kelvin}}
}
\newglossaryentry{constPressureSpecificHeat}{
    type=latin,
    name={\ensuremath{\gls{specificHeat}_{\gls{pressure}}}},
    description={Constant-pressure specific heat of a pure substance},
    unit={\si{\joule\per\kilo\gram\per\kelvin}}
}
\newglossaryentry{specificHeatRatio}{
    type=latin,
    name={\ensuremath{k}},
    description={Specific heat ratio},
    unit={\si{-}}
}
\newglossaryentry{fugacity}{
    type=latin,
    name={\ensuremath{f}},
    description={Fugacity of a system},
    unit={\si{\pascal}}
}
\newglossaryentry{activity}{
    type=latin,
    name={\ensuremath{a}},
    description={Activity of a component in a mixture},
    unit={-}
}
\newglossaryentry{volumeExpansivity}{
    type=greek,
    name={\ensuremath{\alpha_{\gls{pressure}}}},
    description={Volume expansivity of a system},
    unit={\si{\per\kelvin}}
}
\newglossaryentry{isothermalCompressibility}{
    type=greek,
    name={\ensuremath{\beta_{\gls{temperature}}}},
    description={Isothermal compressibility of a system},
    unit={\si{\per\pascal}}
}
\newglossaryentry{volumetricThermalExpansionCoeff}{
    type=greek,
    name={\ensuremath{\beta}},
    description={Volumetric thermal expansion coefficient of a fluid},
    unit={\si{\per\kelvin}}
}
\newglossaryentry{latentHeat}{
    type=latin,
    name={\ensuremath{\gls{intEnthalpy}_{fg}}},
    description={Latent heat},
    unit={\si{\kilo\joule\per\kilo\gram}}
}
\newglossaryentry{adiabaticCompressibility}{
    type=greek,
    name={\ensuremath{\beta_{\gls{entropy}}}},
    description={Adiabatic compressibility of a system},
    unit={\si{\per\pascal}}
}
\newglossaryentry{moleFraction}{
    type=latin,
    name={\ensuremath{y}},
    description={%
        Mole fraction of a component in a mixture
    },
    unit={-}
}
\newglossaryentry{mixture}{
    type=subscript,
    name={\ensuremath{m}},
    description={%
        Indicates a proprty of a homogeneous mixture.
    }
}
\newglossaryentry{reactionExtent}{
    type=greek,
    name={\ensuremath{\varepsilon}},
    description={Extent of a chemical reaction (or degree of advancement)},
    unit={\si{\mol}}
}
\newglossaryentry{formationEnthalpy}{
    type=greek,
    name={\ensuremath{\molar{\gls{intEnthalpy}}^{\mathrm{0}}_{f}}},
    description={Enthalpy of formation of a component},
    unit={\si{\kilo\joule\per\kilo\mol}}
}
\newglossaryentry{formationGibbsFreeEnergy}{
    type=greek,
    name={\ensuremath{\molar{\gls{intGibbsFreeEnergy}}^{\mathrm{0}}_{f}}},
    description={Gibbs free energy of formation of a component},
    unit={\si{\kilo\joule\per\kilo\mol}}
}
\newglossaryentry{absoluteEntropy}{
    type=greek,
    name={\ensuremath{\molar{\gls{intEntropy}}^{\mathrm{0}}}},
    description={Absolute entropy of a substance},
    unit={\si{\kilo\joule\per\kilo\mol\per\kelvin}}
}
\newglossaryentry{standardState}{
    type=superscript,
    name={\ensuremath{\mathrm{0}}},
    description={Indicates the standard state}
}
\newglossaryentry{standardTemperature}{
    type=latin,
    name={\ensuremath{\gsuper{temperature}{standardState}}},
    description={Standard temperature, \SI{298.15}{\kelvin}},
    unit={\si{\kelvin}}
}
\newglossaryentry{standardPressure}{
    type=latin,
    name={\ensuremath{\gsuper{pressure}{standardState}}},
    description={Standard pressure, \SI{0.1}{\mega\pascal}},
    unit={\si{\mega\pascal}}
}
\newglossaryentry{carbonQuantity}{
    type=hidden,
    name={\ensuremath{\alpha}},
    description={%
        Quantity of carbon atoms in the generic
        $\mathrm{C}_{\alpha}\mathrm{H}_{\beta}\mathrm{O}_{\gamma}\mathrm{N}_{\delta}$
    },
    unit={-}
}
\newglossaryentry{hidrogenQuantity}{
    type=hidden,
    name={\ensuremath{\beta}},
    description={%
        Quantity of hidrogen atoms in the generic
        $\mathrm{C}_{\alpha}\mathrm{H}_{\beta}\mathrm{O}_{\gamma}\mathrm{N}_{\delta}$
    },
    unit={-}
}
\newglossaryentry{oxygenQuantity}{
    type=hidden,
    name={\ensuremath{\gamma}},
    description={%
        Quantity of oxigen atoms in the generic
        $\mathrm{C}_{\alpha}\mathrm{H}_{\beta}\mathrm{O}_{\gamma}\mathrm{N}_{\delta}$
    },
    unit={-}
}
\newglossaryentry{nitrogenQuantity}{
    type=hidden,
    name={\ensuremath{\delta}},
    description={%
        Quantity of nitrogen atoms in the generic
        $\mathrm{C}_{\alpha}\mathrm{H}_{\beta}\mathrm{O}_{\gamma}\mathrm{N}_{\delta}$
    },
    unit={-}
}
\newglossaryentry{chemicalGenericFuel}{
    type=hidden,
    name={\ensuremath{%
        \mathrm{C}_{\alpha}\mathrm{H}_{\beta}\mathrm{O}_{\gamma}\mathrm{N}_{\delta}%
    }},
    description={%
        Chemical formula for a generic fuel
    },
    unit={-}
}
\newglossaryentry{airFuelRatio}{
    type=latin,
    name={\ensuremath{\mathit{AC}}},
    description={Air-fuel ratio},
    unit={-}
}
\newglossaryentry{stoichiometricAirFuelRatio}{
    type=latin,
    name={\ensuremath{\mathit{AC_s}}},
    description={Stoichiometric or theoretical air-fuel ratio},
    unit={-}
}
\newglossaryentry{molarAirFuelRatio}{
    type=latin,
    name={\ensuremath{a_s}},
    description={Stoichiometric molar air-fuel ratio},
    unit={-}
}
\newglossaryentry{fuelAirRatio}{
    type=latin,
    name={\ensuremath{\mathit{CA}}},
    description={Fuel-air ratio},
    unit={-}
}
\newglossaryentry{stoichiometricFuelAirRatio}{
    type=latin,
    name={\ensuremath{\mathit{CA_s}}},
    description={Stoichiometric or theoretical fuel-air ratio},
    unit={-}
}
\newglossaryentry{equivalenceRatio}{
    type=latin,
    name={\ensuremath{\phi}},
    description={Air-fuel equivalent ratio},
    unit={-}
}
\newglossaryentry{rEquivalenceRatio}{
    type=latin,
    name={\ensuremath{\lambda}},
    description={Reciprocal of the air-fuel equivalent ratio},
    unit={-}
}
\newglossaryentry{chProducts}{
    type=subscript,
    name={\ensuremath{P}},
    description={Indicates the products of a chemical reaction}
}
\newglossaryentry{chReactants}{
    type=subscript,
    name={\ensuremath{R}},
    description={Indicates the reactants of a chemical reaction}
}
\newglossaryentry{flameAdiabaticTemperature}{
    type=latin,
    name={\ensuremath{\gls{temperature}_{ad}}},
    description={Flame adiabatic temperature},
    unit={\si{\kelvin}}
}
\newglossaryentry{combustionEnthalpy}{
    type=greek,
    name={\ensuremath{%
        \molar{\gls{intEnthalpy}}^{\mathrm{0}}_{\gls{chReactants}\gls{chProducts}}}%
    },
    description={Enthalpy of combustion of a fuel},
    unit={\si{\kilo\joule\per\kilo\mol}}
}
\newglossaryentry{LagrangeMultiplyer}{
    type=greek,
    name={\ensuremath{\lambda}},
    description={Lagrange multiplyer},
    unit={-}
}
\newglossaryentry{chEquilibriumConstant}{
    type=latin,
    name={\ensuremath{K}},
    description={Constant of equilibrium of a chemical reaction},
    unit={-}
}
\newglossaryentry{PoyntingFactor}{
    type=latin,
    name={\ensuremath{F}},
    description={Poynting factor},
    unit={-}
}

\newglossaryentry{molecularMass}{
    type=latin,
    name={\ensuremath{M}},
    description={Molecular mass},
    unit={\si{\kilo\mol}}
}

\newglossaryentry{convectionHeatTransferCoeff}{
    type=latin,
    name={\ensuremath{h}},
    description={Convection heat transfer coefficient},
    unit={\si{\watt\per\square\meter\kelvin}}
}
\newglossaryentry{meanConvectionHeatTransferCoeff}{
    type=hidden,
    name={\ensuremath{\avg{h}}},
    description={Mean convection heat transfer coefficient},
    unit={\si{\watt\per\square\meter\kelvin}}
}
\newglossaryentry{convectionMassTransferCoeff}{
    type=latin,
    name={\ensuremath{h_m}},
    description={Convection mass transfer coefficient},
    unit={\si{\meter\per\second}}
}
\newglossaryentry{meanConvectionMassTransferCoeff}{
    type=hidden,
    name={\ensuremath{\avg{h}_m}},
    description={Mean convection mass transfer coefficient},
    unit={\si{\meter\per\second}}
}
\newglossaryentry{conductionShapeFactor}{
    type=latin,
    name={\ensuremath{S}},
    description={Conduction shape factor},
    unit={\si{\meter}}
}
\newglossaryentry{freeStream}{
    type=subscript,
    name={\ensuremath{\infty}},
    description={Indicates free-stream properties}
}
\newglossaryentry{velocityLayerThickness}{
    type=latin,
    name={\ensuremath{\delta}},
    description={Thickness of the velocity boundary layer},
    unit={\si{\meter}}
}
\newglossaryentry{thermalLayerThickness}{
    type=latin,
    name={\ensuremath{\delta_t}},
    description={Thickness of the thermal boundary layer},
    unit={\si{\meter}}
}
\newglossaryentry{totalHeatTransferRate}{
    type=latin,
    name={\ensuremath{q}},
    description={%
        Total heat transfer rate
    },
    unit={\si{\watt}}
}
\newglossaryentry{thermalResistance}{
    type=latin,
    name={\ensuremath{R_t}},
    description={%
        Thermal resistance
    },
    unit={\si{\kelvin\per\watt}}
}
\newglossaryentry{convectiveThermalResistance}{
    type=hidden,
    name={\ensuremath{R_{t,\mathrm{conv}}}},
    description={%
        Convective thermal resistance
    },
    unit={\si{\kelvin\per\watt}}
}
\newglossaryentry{conductionThermalResistance}{
    type=hidden,
    name={\ensuremath{R_{t,\mathrm{cond}}}},
    description={%
        Thermal resistance for condution
    },
    unit={\si{\kelvin\per\watt}}
}
\newglossaryentry{radiationThermalResistance}{
    type=hidden,
    name={\ensuremath{R_{t,\mathrm{rad}}}},
    description={%
        Thermal resistance for radiation
    },
    unit={\si{\kelvin\per\watt}}
}
\newglossaryentry{totalThermalResistance}{
    type=hidden,
    name={\ensuremath{R_{\mathrm{tot}}}},
    description={%
        Total thermal resistance of a thermal circuit
    },
    unit={\si{\kelvin\per\watt}}
}
\newglossaryentry{globalHeatTransferCoeff}{
    type=hidden,
    name={\ensuremath{U}},
    description={%
        Global heat transfer coefficient
    },
    unit={\si{\watt\per\square\meter\kelvin}}
}

\newglossaryentry{electricalResistance}{
    type=hidden,
    name={\ensuremath{R_e}},
    description={%
        Electrical resistance
    },
    unit={\si{\ohm}}
}
\newglossaryentry{electricalCurrent}{
    type=hidden,
    name={\ensuremath{I}},
    description={%
        Electrical current
    },
    unit={\si{\ampere}}
}

\newglossaryentry{radiation}{
    type=subscript,
    name={\ensuremath{\mathrm{rad}}},
    description={Indicates radiation heat exchange}
}
\newglossaryentry{surroundings}{
    type=subscript,
    name={\ensuremath{\mathrm{viz}}},
    description={Indicates the properties of the surroundings surfaces}
}
\newglossaryentry{emissivePower}{
    type=hidden,
    name={\ensuremath{E}},
    description={%
        Emissive power of a surface
    },
    unit={\si{\watt\per\square\meter}}
}
\newglossaryentry{irradiation}{
    type=hidden,
    name={\ensuremath{G}},
    description={%
        Irradiation on a surface
    },
    unit={\si{\watt\per\square\meter}}
}
\newglossaryentry{StefanBoltzmannConstant}{
    type=hidden,
    name={\ensuremath{\sigma}},
    description={%
        Stefan-Boltazmann constant ($ =
        \SI{5.67e-8}{\watt\per\square\meter\kelvin^4}$)
    },
    unit={\si{\watt\per\square\meter\kelvin^4}}
}
\newglossaryentry{emissivity}{
    type=hidden,
    name={\ensuremath{\varepsilon}},
    description={%
        Emissivity of a surface
    },
    unit={\si{-}}
}
\newglossaryentry{absorptivity}{
    type=hidden,
    name={\ensuremath{\alpha}},
    description={%
        Absorptivity of a surface
    },
    unit={\si{-}}
}

\newglossaryentry{cvNumber}{
    type=latin,
    name={\ensuremath{N_{CV}}},
    description={Number of control volumes in a finite volume mesh},
    unit={\si{-}}
}

\newglossaryentry{centroid}{
    type=subscript,
    name={\ensuremath{C}},
    description={%
        Indicates variables evaluated at the centroid of a control volume
    }
}
\newglossaryentry{nbCentroid}{
    type=subscript,
    name={\ensuremath{N}},
    description={%
        Indicates variables evaluated at the centroid of a neighbor control
        volume to cell $V_C$
    }
}
\newglossaryentry{owner}{
    type=subscript,
    name={\ensuremath{\gls{centroid}}},
    description={%
        Indicates variables evaluated at the centroid of a the face owner
        control volume
    }
}
\newglossaryentry{neighbor}{
    type=subscript,
    name={\ensuremath{\gls{nbCentroid}}},
    description={%
        Indicates variables evaluated at the centroid of a face neighbor
        control volume
    }
}
\newglossaryentry{face}{
    type=subscript,
    name={\ensuremath{f}},
    description={%
        Indicate a face of a polyhedral control volume and a property evaluated
        at it
    }
}
\newglossaryentry{skewFace}{
    type=subscript,
    name={\ensuremath{\gls{face}^{\prime}}},
    description={%
        Indicates the point where the vector joining two face adjacent cell's
        centroids intercepts the face
    }
}

\newglossaryentry{fvCellVolume}{
    type=latin,
    name={\ensuremath{\gls{volume}_{\gls{centroid}}}},
    description={Volume of a cell with centroid $C$ in a finite volume mesh},
    unit={\si{\cubic\meter}}
}
\newglossaryentry{fvCellSurface}{
    type=latin,
    name={\ensuremath{\gls{surface}_{\gls{centroid}}}},
    description={Surface of a cell with centroid $C$ in a finite volume mesh},
    unit={\si{\square\meter}}
}
\newglossaryentry{fvFaceNormal}{
    type=latin,
    name={\ensuremath{\vec{S}_{\gls{face}}}},
    description={Normal vector to the face $f$ of a polyhedral cell},
    unit={\si{\square\meter}}
}
\newglossaryentry{joinCentroidVector}{
    type=latin,
    name={\ensuremath{\vec{d}_{\gls{centroid}\gls{nbCentroid}}}},
    description={%
        Vector joining the centroids of adjacent cells in a finite volume mesh
    },
    unit={\si{\meter}}
}
\newglossaryentry{skewnessVector}{
    type=latin,
    name={\ensuremath{\vec{m}}},
    description={%
        Vector joining the face centroid with the interception point between
        the face and the line between the owner and neighbour cells.
    },
    unit={\si{\meter}}
}
\newglossaryentry{facePyramidHeigth}{
    type=latin,
    name={\ensuremath{\vec{h}_{\gls{face}\gls{centroid}}}},
    description={%
        Heigth of a cell face pyramid -- is the opposite of the vector joining
        the cell centroid and the face centroid
    },
    unit={\si{\meter}}
}
\newglossaryentry{centroidFaceVector}{
    type=latin,
    name={\ensuremath{\vec{d}_{\gls{centroid}\gls{face}}}},
    description={%
        Vector joining the centroids of a cell \gls{fvCellVolume} and the
        centroid of a face \gls{face}
    },
    unit={\si{\meter}}
}
\newglossaryentry{ownerFaceVector}{
    type=latin,
    name={\ensuremath{\vec{d}_{\gls{owner}\gls{face}}}},
    description={%
        Vector joining the centroid of a face owner cell and the centroid of
        the face \gls{face}
    },
    unit={\si{\meter}}
}
\newglossaryentry{neighFaceVector}{
    type=latin,
    name={\ensuremath{\vec{d}_{\gls{neighbor}\gls{face}}}},
    description={%
        Vector joining the centroid of a face neighbor cell and the centroid of
        the face \gls{face}
    },
    unit={\si{\meter}}
}

\newglossaryentry{projCentroidNormalVec}{
    type=hidden,
    name={\ensuremath{\vec{d}_{fn}}},
    description={%
        Vector formed by the projection of \gls{joinCentroidVector}
        on \gls{sNormalVector}
    },
    unit={\si{\meter}}
}
\newglossaryentry{projCentroidNormalNorm}{
    type=latin,
    name={\ensuremath{d_{fn}}},
    description={%
        Projection of the vector \gls{joinCentroidVector} on the face normal
        \gls{sNormalVector}
    },
    unit={\si{\meter}}
}
\newglossaryentry{advWeightFactor}{
    type=hidden,
    name={\ensuremath{w}},
    description={%
        Weight of cell \gls{centroid} on the advective interpolation profile
    },
    unit={\si{-}}
}
\newglossaryentry{faceVolumeSwept}{
    type=latin,
    name={\ensuremath{\dot{V}_f}},
    description={%
        Volume swept by face \gls{face} of control volume \gls{fvCellVolume}
    },
    unit={\si{\cubic\meter\per\second}}
}
\newglossaryentry{normalDecomposition1}{
    type=latin,
    name={\ensuremath{\vec{\beta}}},
    description={%
        Vector joining the centroid of face \gls{face} and the neighbor cell
        centroid \gls{nbCentroid}
    },
    unit={\si{-}}
}
\newglossaryentry{normalDecomposition1Norm}{
    type=hidden,
    name={\ensuremath{\beta}},
    description={%
        Modulus of the vector joining the centroid of face \gls{face} and the
        neighbor cell centroid \gls{nbCentroid}
    },
    unit={\si{-}}
}
\newglossaryentry{normalDecomposition2}{
    type=greek,
    name={\ensuremath{\vec{\kappa}}},
        description={Complement vector of the decomposition of
        \gls{fvFaceNormal} in \gls{normalDecomposition1}},
        unit={\si{-}}
}
\newglossaryentry{fvInterpolationWeight}{
    type=hidden,
    name={\ensuremath{g}},
    description={%
        Generic symbol for the interpolation weights of the finite volume
        method
    },
    unit={\si{-}}
}
\newglossaryentry{joinCentroidCellFace}{
    type=latin,
    name={\ensuremath{\vec{d}_{\gls{face}\gls{centroid}}}},
    description={%
        Vector ligando os centroides \gls{centroid} da célula e de uma face
        \gls{face}
    },
    unit={\si{\meter}}
}
\newglossaryentry{joinNbCentroidCellFace}{
    type=latin,
    name={\ensuremath{\vec{d}_{\gls{face}\gls{nbCentroid}}}},
    description={%
        Vector ligando os centroides da célula vizinha \gls{nbCentroid} da
        célula e de uma face \gls{face}
    },
    unit={\si{\meter}}
}

\newglossaryentry{transportVariable}{
    type=greek,
    name={\ensuremath{\phi}},
    description={%
        Generic intensive (per unit mass) variable transported by a fluid flow
    },
    unit={$[\phi]/$\si{\kilo\gram}}
}
\newglossaryentry{integralTransportVariable}{
    type=greek,
    name={\ensuremath{\Phi}},
    description={%
        Generic intensive variable transported by a fluid flow integrated in a
        finite control volume
    },
    unit={$[\Phi]$}
}
\newglossaryentry{transportVariablePerVolume}{
    type=greek,
    name={\ensuremath{\varphi}},
    description={%
        Generic intensive (per unit volume) variable transported by a fluid
        flow
    },
    unit={$[\Phi]/$\si{\cubic\meter}}
}
\newglossaryentry{transportDiffusivity}{
    type=greek,
    name={\ensuremath{\Gamma^{\phi}}},
    description={Diffusivity of \gls{transportVariable}},
    unit={\si{\square\meter\per\second}}
}
\newglossaryentry{transportSource}{
    type=latin,
    name={\ensuremath{S^{\phi}}},
    description={%
        Source and/or sink terms of the differential transport equation of
        \gls{transportVariable}
    },
    unit={$[\phi]/$\si{\cubic\meter\per\second}}
}
\newglossaryentry{linearTransportSource}{
    type=latin,
    name={\ensuremath{\bar{S}^{\phi}}},
    description={%
        Mean source and/or sink terms of the integral transport equation of
        \gls{transportVariable}
    },
    unit={$[\phi]/$\si{\cubic\meter\per\second}}
}
\newglossaryentry{fvCoeffs}{
    type=latin,
    name={\ensuremath{a}},
    description={%
        Generic coefficients of the linearized transport equation for the
        volume $V_C$
    },
    unit={\si{\kilo\gram\per\second}}
}
\newglossaryentry{fvDiagCoeffs}{
    type=latin,
    name={\ensuremath{\gls{fvCoeffs}_{\gls{centroid}}}},
    description={%
        Diagonal coefficients of the linearized transport equation for the
        volume $V_C$
    },
    unit={\si{\kilo\gram\per\second}}
}
\newglossaryentry{fvOffDiagCoeffs}{
    type=latin,
    name={\ensuremath{\gls{fvCoeffs}_{\gls{nbCentroid}}}},
    description={%
        Off-diagonal coefficients of the linearized transport equation for the
        volume $V_C$
    },
    unit={\si{\kilo\gram\per\second}}
}
\newglossaryentry{fvMatrix}{
    type=latin,
    name={\ensuremath{\matrix{A}}},
    description={%
        Coefficients matrix of a system of equations assembled for a finite
        volume mesh
    },
    unit={$[\phi]$}
}
\newglossaryentry{sourceScalar}{
    type=hidden,
    name={\ensuremath{b}},
    description={%
        Source term of the linearized transport equation for a control volume
        $V_C$
    },
    unit={$[\phi]$\si{\kilo\gram\per\second}}
}
\newglossaryentry{fvSource}{
    type=hidden,
    name={\ensuremath{\cvec{b}}},
    description={%
        Source term of the linearized system equation for a finite volume mesh
    },
    unit={\si{-}}
}
\newglossaryentry{relaxationFactor}{
    type=hidden,
    name={\ensuremath{\alpha}},
    description={%
        Relaxation factor
    },
    unit={\si{-}}
}
\newglossaryentry{sourceVector}{
    type=latin,
    name={\ensuremath{\vec{b}}},
    description={%
        Source vector of the linearized momentum equations for a control volume
    },
    unit={$[\phi]$\si{\kilo\gram\per\second}}
}
\newglossaryentry{pressureCorrection}{
    type=latin,
    name={\ensuremath{p'}},
    description={%
        Pressure correction field in the pressure-velocity algorithms to solve
        the coupled Navier-Stokes equations
    },
    unit={\si{\pascal}}
}
\newglossaryentry{transportOperator}{
    type=hidden,
    name={\ensuremath{\matrix{H}}},
    description={Transport operator},
    unit={\si{-}}
}
\newglossaryentry{sourceOperator}{
    type=hidden,
    name={\ensuremath{\matrix{B}}},
    description={Source operator},
    unit={\si{-}}
}
\newglossaryentry{pressureDiffOperator}{
    type=hidden,
    name={\ensuremath{\matrix{D}}},
    description={Pressure diffusive operator},
    unit={\si{-}}
}

\newglossaryentry{meshNonOrthog}{
    type=latin,
    name={\ensuremath{mesh_{nonOrtho}}},
    description={Mesh non-orthogonality}
}

\newglossaryentry{meshSkewness}{
    type=latin,
    name={\ensuremath{mesh_{skew}}},
    description={Mesh skewness}
}

\newglossaryentry{faceTwist}{
    type=latin,
    name={\ensuremath{mesh_{faceTwist}}},
    description={Mesh face twist}
}

\newglossaryentry{faceWeight}{
    type=latin,
    name={\ensuremath{mesh_{faceWeight}}},
    description={Mesh face weight}
}

\newglossaryentry{faceVolumeRatio}{
    type=latin,
    name={\ensuremath{mesh_{faceVolRatio}}},
    description={Mesh face volume ratio}
}

\newglossaryentry{faceTriangTwist}{
    type=latin,
    name={\ensuremath{mesh_{faceTriTwist}}},
    description={Mesh face twist}
}

\newglossaryentry{cellConvexity}{
    type=latin,
    name={\ensuremath{mesh_{cellConcave}}},
    description={Mesh cells convexity: it measures if a cell is concave.}
}

\newglossaryentry{pyramidVolume}{
    type=latin,
    name={\ensuremath{mesh_{cellPyrVol}}},
    description={Mesh cells pyramids volume}
}

\newglossaryentry{cellTetQuality}{
    type=latin,
    name={\ensuremath{mesh_{tetQuality}}},
    description={Mesh cells tetrahedral decomposition quality}
}
\newglossaryentry{polyCellTensor}{
    type=latin,
    name={\ensuremath{\tensor{T}}},
    description={%
        Second-order tensor formed by the outer product of the face vector of a
        generic polyhedral cell
    }
}
\newglossaryentry{cellDeterminant}{
    type=latin,
    name={\ensuremath{mesh_{cellDeterminant}}},
    description={Mesh face twist}
}

\newglossaryentry{dynamicViscosity}{
    type=greek,
    name={\ensuremath{\mu}},
    description={Dynamic viscosity of a fluid},
    unit={\si{\pascal\second}}
}
\newglossaryentry{surfaceTension}{
    type=greek,
    name={\ensuremath{\sigma}},
    description={Surface tension},
    unit={\si{\newton\per\meter}}
}
\newglossaryentry{secondViscosity}{
    type=greek,
    name={\ensuremath{\lambda}},
    description={%
        Second viscosity of a fluid (sometimes referred to as
        \emph{bulk viscosity})
    },
    unit={\si{\pascal\second}}
}
\newglossaryentry{kinematicViscosity}{
    type=greek,
    name={\ensuremath{\nu}},
    description={Kinematic viscosity of a fluid},
    unit={\si{\square\meter\per\second}}
}
\newglossaryentry{pressure}{
    type=latin,
    name={\ensuremath{p}},
    description={Pressure field of a fluid flow},
    unit={\si{\pascal}}
}
\newglossaryentry{viscousDissipationRateFunction}{
    type=greek,
    name={\ensuremath{\Phi}},
    description={Viscous dissipation rate function},
    unit={\si{\watt\per\cubic\meter}}
}
\newglossaryentry{meshDiffusivity}{
    type=greek,
    name={\ensuremath{\gamma}},
    description={%
        Mesh diffusivity for the Laplace equation governing the mesh motion
    },
    unit={\si{-}}
}
\newglossaryentry{centroidToBoundaryDistance}{
    type=latin,
    name={\ensuremath{l}},
    description={Cell centroid distance to a selected boundary},
    unit={\si{\meter}}
}
\newglossaryentry{viscousCauchyStress}{
    type=greek,
    name={\ensuremath{\gvec{\tau}}},
    description={Viscous part of the Cauchy stress tensor of a fluid flow},
    unit={\si{\pascal}}
}
\newglossaryentry{viscousCauchyStressComp}{
    type=hidden,
    name={\ensuremath{\tau}},
    description={Viscous part of the Cauchy stress tensor of a fluid flow},
    unit={\si{\pascal}}
}
\newglossaryentry{vorticityTensor}{
    type=greek,
    name={\ensuremath{\tensor{\Omega}}},
    description={Vorticity or spin tensor},
    unit={\si{\per\second}}
}
\newglossaryentry{vorticityTensorComp}{
    type=hidden,
    name={\ensuremath{\Omega}},
    description={Vorticity or spin tensor component},
    unit={\si{\per\second}}
}
\newglossaryentry{vorticity}{
    type=greek,
    name={\ensuremath{\gvec{\zeta}}},
    description={Vorticity vector},
    unit={\si{\per\second}}
}
\newglossaryentry{vorticityComp}{
    type=hidden,
    name={\ensuremath{\zeta}},
    description={Vorticity vector component},
    unit={\si{\per\second}}
}
\newglossaryentry{circulation}{
    type=hidden,
    name={\ensuremath{\Gamma}},
    description={Circulation},
    unit={\si{\square\meter\per\second}}
}
\newglossaryentry{ReynoldsNumber}{
    type=latin,
    name={\ensuremath{Re}},
    description={Reynolds number},
    unit={\si{-}}
}
\newglossaryentry{PecletNumber}{
    type=latin,
    name={\ensuremath{Pe}},
    description={Péclet number},
    unit={\si{-}}
}
\newglossaryentry{frictionCoefficient}{
    type=latin,
    name={\ensuremath{C_f}},
    description={Friction coefficient},
    unit={\si{-}}
}
\newglossaryentry{frictionFactor}{
    type=latin,
    name={\ensuremath{f}},
    description={Friction factor for head loss computation in internal flow},
    unit={\si{-}}
}
\newglossaryentry{MachNumber}{
    type=latin,
    name={\ensuremath{Ma}},
    description={Mach number},
    unit={\si{-}}
}
\newglossaryentry{soundSpeed}{
    type=latin,
    name={\ensuremath{c}},
    description={Sound speed},
    unit={\si{\meter\per\second}}
}
\newglossaryentry{WeberNumber}{
    type=latin,
    name={\ensuremath{We}},
    description={Weber number},
    unit={\si{-}}
}
\newglossaryentry{FroudeNumber}{
    type=latin,
    name={\ensuremath{Fr}},
    description={Froude number},
    unit={\si{-}}
}
\newglossaryentry{EulerNumber}{
    type=latin,
    name={\ensuremath{Eu}},
    description={Euler number},
    unit={\si{-}}
}
\newglossaryentry{CavitationNumber}{
    type=latin,
    name={\ensuremath{Ca}},
    description={Cavitation number},
    unit={\si{-}}
}
\newglossaryentry{CourantNumber}{
    type=latin,
    name={\ensuremath{Co}},
    description={Courant number},
    unit={\si{-}}
}
\newglossaryentry{PrandtlNumber}{
    type=latin,
    name={\ensuremath{Pr}},
    description={%
        Prandtl number
    },
    unit={\si{-}}
}
\newglossaryentry{SchmidtNumber}{
    type=latin,
    name={\ensuremath{Sc}},
    description={%
        Schmidt number
    },
    unit={\si{-}}
}
\newglossaryentry{NusseltNumber}{
    type=latin,
    name={\ensuremath{Nu}},
    description={%
        Nusselt number
    },
    unit={\si{-}}
}
\newglossaryentry{SherwoodNumber}{
    type=latin,
    name={\ensuremath{Sh}},
    description={%
        Sherwood number
    },
    unit={\si{-}}
}
\newglossaryentry{BiotNumber}{
    type=latin,
    name={\ensuremath{Bi}},
    description={Biot number},
    unit={\si{-}}
}
\newglossaryentry{GrashofNumber}{
    type=latin,
    name={\ensuremath{Gr}},
    description={Grashof number},
    unit={\si{-}}
}
\newglossaryentry{RayleighNumber}{
    type=latin,
    name={\ensuremath{Ra}},
    description={Rayleigh number},
    unit={\si{-}}
}
\newglossaryentry{FourierNumber}{
    type=latin,
    name={\ensuremath{Fo}},
    description={Fourier number},
    unit={\si{-}}
}
\newglossaryentry{shearRate}{
    type=latin,
    name={\ensuremath{\dot{\gamma}}},
    description={%
        Shear rate components in a fluid flow
    },
    unit={\si{\per\second}}
}
\newglossaryentry{yieldStress}{
    type=hidden,
    name={\ensuremath{\tau_C}},
    description={Yield stress of a Herschel-Bulkley fluid},
    unit={\si{\pascal}}
}
\newglossaryentry{consistency}{
    type=hidden,
    name={\ensuremath{k_n}},
    description={Consistency of a Herschel-Bulkley fluid},
    unit={\si{\pascal\second^n}}
}
\newglossaryentry{indexFlow}{
    type=hidden,
    name={\ensuremath{n}},
    description={Index flow of a Herschel-Bulkley fluid},
    unit={\si{-}}
}
\newglossaryentry{apparentViscosity}{
    type=greek,
    name={\ensuremath{\eta^{\gls{fluid}}}},
    description={Apparent viscosity of a non-Newtonian fluid},
    unit={\si{\pascal\second}}
}
\newglossaryentry{CassonCoefficient}{
    type=latin,
    name={\ensuremath{k}},
    description={Consistency coefficient of Casson's model},
    unit={\si{\pascal\second}}
}

\newglossaryentry{firstInvariant}{
    type=latin,
    name={\ensuremath{I_1}},
    description={First principal invariant of the deformation tensors},
    unit={\si{-}}
}
\newglossaryentry{secondInvariant}{
    type=latin,
    name={\ensuremath{I_2}},
    description={Second principal invariant of the deformation tensors},
    unit={\si{-}}
}
\newglossaryentry{thirdInvariant}{
    type=latin,
    name={\ensuremath{I_3}},
    description={Third principal invariant of the deformation tensors},
    unit={\si{-}}
}
\newglossaryentry{modifiedSecondInvariant}{
    type=latin,
    name={\ensuremath{J_2}},
    description={Modified second invariant of the rate of deformation tensor},
    unit={\si{-}}
}
\newglossaryentry{elasticModuliTensor}{
    type=latin,
    name={\ensuremath{\fourthTensor{C}}},
    description={Elastic moduli tensor},
    unit={\si{\pascal}}
}
\newglossaryentry{elasticModuliTensorComp}{
    type=hidden,
    name={\ensuremath{C}},
    description={Elastic moduli tensor component},
    unit={\si{\pascal}}
}
\newglossaryentry{YoungModulus}{
    type=latin,
    name={\ensuremath{E}},
    description={Young's modulus of a solid},
    unit={\si{\pascal}}
}
\newglossaryentry{shearModulus}{
    type=latin,
    name={\ensuremath{G}},
    description={Shear modulus of a solid},
    unit={\si{\pascal}}
}
\newglossaryentry{OgdenCoeffs}{
    type=greek,
    name={\ensuremath{\mu}},
    description={Generic representation of Ogden's law material constants},
    unit={\si{\pascal}}
}
\newglossaryentry{PoissonRatio}{
    type=greek,
    name={\ensuremath{\nu^{\gls{solid}}}},
    description={Poisson's ratio of a solid},
    unit={\si{-}}
}
\newglossaryentry{plasticYieldStress}{
    type=greek,
    name={\ensuremath{\sigma_Y}},
    description={Yield stress for plastic deformation of a solid},
    unit={\si{\pascal}}
}
\newglossaryentry{strainEnergyFunction}{
    type=greek,
    name={\ensuremath{\Psi}},
    description={%
        Deformation work function or strain energy function
    },
    unit={\si{\joule\per\cubic\meter}}
}
\newglossaryentry{deformationWorkFunction}{
    type=latin,
    name={\ensuremath{\Psi}},
    description={%
        Deformation work function or strain energy function -- function of
        \gls{rightCauchyGreenDeformation}. Energy from deformation work per
        unit volume
    },
    unit={\si{\joule\per\cubic\meter}}
}
\newglossaryentry{lameConst1}{
    type=greek,
    name={\ensuremath{\mu^{\gls{solid}}}},
    description={First Lamé's constant of a solid},
    unit={\si{\pascal}}
}
\newglossaryentry{lameConst2}{
    type=greek,
    name={\ensuremath{\lambda^{\gls{solid}}}},
    description={Second Lamé's constant of a solid},
    unit={\si{\pascal}}
}
\newglossaryentry{nominalStress}{
    type=latin,
    name={\ensuremath{\tensor{P}}},
    description={Nominal stress tensor},
    unit={\si{\pascal}}
}
\newglossaryentry{nominalStressComp}{
    type=hidden,
    name={\ensuremath{P}},
    description={Nominal stress tensor component},
    unit={\si{\pascal}}
}
\newglossaryentry{strainRateTensor}{
    type=latin,
    name={\ensuremath{\tensor{D}}},
    description={Rate of deformation tensor},
    unit={\si{\per\second   }}
}
\newglossaryentry{strainRateTensorComp}{
    type=hidden,
    name={\ensuremath{D}},
    description={Rate of deformation tensor component},
    unit={\si{\per\second}}
}
\newglossaryentry{deformationGradient}{
    type=latin,
    name={\ensuremath{\tensor{F}}},
    description={Deformation gradient tensor},
    unit={\si{-}}
}
\newglossaryentry{elastic}{
    type=superscript,
    name={\ensuremath{e}},
    description={Superscript indicating elastic part of a deformation},
    unit={\si{-}}
}
\newglossaryentry{plastic}{
    type=superscript,
    name={\ensuremath{p}},
    description={Superscript indicating plastic part of a deformation},
    unit={\si{-}}
}
\newglossaryentry{deformationGradientComp}{
    type=hidden,
    name={\ensuremath{F}},
    description={Deformation gradient tensor component},
    unit={\si{-}}
}
\newglossaryentry{rotationTensor}{
    type=latin,
    name={\ensuremath{\tensor{R}}},
    description={Rotation tensor},
    unit={\si{-}}
}

\newglossaryentry{stretch}{
    type=greek,
    name={\ensuremath{\lambda}},
    description={%
        Uniaxial stretch of a solid deforming
    },
    unit={\si{-}}
}
\newglossaryentry{deformationGradComp}{
    type=hidden,
    name={\ensuremath{F}},
    description={Deformation gradient tensor component},
    unit={\si{-}}
}
\newglossaryentry{solidDisplacement}{
    type=latin,
    name={\ensuremath{\vec{u}}},
    description={Solid displacement vector},
    unit={m}
}
\newglossaryentry{solidDisplacementComp}{
    type=hidden,
    name={\ensuremath{u}},
    description={Solid displacement vector component},
    unit={\si{-}}
}
\newglossaryentry{xSolidDisplacement}{
    type=hidden,
    name={\ensuremath{u_{\gls{xAxis}}}},
    description={%
        Solid displacement vector component in the x direction of Cartesian
        coordinates
    },
    unit={\si{-}}
}
\newglossaryentry{ySolidDisplacement}{
    type=hidden,
    name={\ensuremath{u_{\gls{yAxis}}}},
    description={%
        Solid displacement vector component in the y direction of Cartesian
        coordinates
    },
    unit={\si{-}}
}
\newglossaryentry{zSolidDisplacement}{
    type=hidden,
    name={\ensuremath{u_{\gls{zAxis}}}},
    description={%
        Solid displacement vector component in the z direction of Cartesian
        coordinates
    },
    unit={\si{-}}
}
\newglossaryentry{secondPiolaKirchhoffStress}{
    type=latin,
    name={\ensuremath{\tensor{S}}},
    description={Second Piola-Kirchhoff stress tensor},
    unit={\si{\pascal}}
}
\newglossaryentry{secondPiolaKirchhoffStressComp}{
    type=hidden,
    name={\ensuremath{S}},
    description={Second Piola-Kirchhoff stress tensor component},
    unit={\si{\pascal}}
}
\newglossaryentry{KirchhoffStress}{
    type=greek,
    name={\ensuremath{\tensor{\sigma}^{\mathrm{K}}}},
    description={Kirchhoff stress tensor},
    unit={\si{\pascal}}
}
\newglossaryentry{KirchhoffStressComp}{
    type=hidden,
    name={\ensuremath{\sigma^{\mathrm{K}}}},
    description={Kirchhoff stress tensor component},
    unit={\si{\pascal}}
}
\newglossaryentry{GLstrain}{
    type=latin,
    name={\ensuremath{\tensor{E}}},
    description={Green-Lagrange strain tensor},
    unit={\si{-}}
}
\newglossaryentry{leftCauchyGreenDeformation}{
    type=latin,
    name={\ensuremath{\tensor{B}}},
    description={Left Cauchy-Green deformation tensor},
    unit={\si{-}}
}
\newglossaryentry{rightCauchyGreenDeformation}{
    type=latin,
    name={\ensuremath{\tensor{C}}},
    description={Right Cauchy-Green deformation tensor},
    unit={\si{-}}
}
\newglossaryentry{2PKstressComp}{
    type=hidden,
    name={\ensuremath{S}},
    description={Second Piola-Kirchhoff stress tensor component},
    unit={\si{\pascal}}
}
\newglossaryentry{GLstrainComp}{
    type=hidden,
    name={\ensuremath{E}},
    description={Green-Lagrange strain tensor component},
    unit={\si{-}}
}
\newglossaryentry{leftCauchyGreenDeformationComp}{
    type=hidden,
    name={\ensuremath{B}},
    description={Left Green-Cauchy deformation tensor component},
    unit={\si{-}}
}
\newglossaryentry{rightCauchyGreenDeformationComp}{
    type=hidden,
    name={\ensuremath{C}},
    description={Right Green-Cauchy deformation tensor component},
    unit={\si{-}}
}
\newglossaryentry{linearStrain}{
    type=greek,
    name={\ensuremath{\tensor{\varepsilon}}},
    description={Linear or engineering strain tensor},
    unit={\si{-}}
}
\newglossaryentry{linearStrainComp}{
    type=hidden,
    name={\ensuremath{\varepsilon}},
    description={Linear or engineering strain tensor},
    unit={\si{-}}
}

\newglossaryentry{trueLinearStrain}{
    type=greek,
    name={\ensuremath{\tensor{e}}},
    description={True or logarithmic strain tensor},
    unit={\si{-}}
}

\newglossaryentry{tensorInvariant}{
    type=hidden,
    name={\ensuremath{I}},
    description={},
    unit={\si{-}}
}
\newglossaryentry{fourthInvariant}{
    type=latin,
    name={\ensuremath{I_4}},
    description={%
        Fourth invariant of the right Cauchy-Green deformation tensor with
        direction information
    },
    unit={\si{-}}
}
\newglossaryentry{fifthInvariant}{
    type=latin,
    name={\ensuremath{I_5}},
    description={%
        Third invariant of the right Cauchy-Green deformation tensor with
        directional information
    },
    unit={\si{-}}
}
\newglossaryentry{sixthInvariant}{
    type=latin,
    name={\ensuremath{I_6}},
    description={%
        Sixth invariant of the right Cauchy-Green deformation tensor with
        directional information
    },
    unit={\si{-}}
}
\newglossaryentry{seventhInvariant}{
    type=latin,
    name={\ensuremath{I_7}},
    description={%
        Seventh invariant of the right Cauchy-Green deformation tensor with
        directional information
    },
    unit={\si{-}}
}
\newglossaryentry{eighthInvariant}{
    type=latin,
    name={\ensuremath{I_8}},
    description={%
        Eighth invariant of the right Cauchy-Green deformation tensor with
        directional information
    },
    unit={\si{-}}
}

\newglossaryentry{MooneyCoeffs}{
    type=hidden,
    name={\ensuremath{c}},
    description={%
        Coefficients of Rivlin series expansion of the strain-energy function
    },
    unit={\si{\kilo\joule\per\cubic\meter}}
}
\newglossaryentry{OgdenExponents}{
    type=hidden,
    name={\ensuremath{\alpha}},
    description={Exponents of Ogden's strain-energy function},
    unit={\si{-}}
}
\newglossaryentry{FungArgumentFunction}{
    type=latin,
    name={\ensuremath{Q}},
    description={%
        Function of the strain tensor components used in the generic Fung-type
        models
    },
    unit={\si{-}}
}
\newglossaryentry{fiberDirection}{
    type=latin,
    name={\ensuremath{\vec{m}}},
    description={%
        Vector of the preferred direction of a fiber reinforced material
    },
    unit={\si{-}}
}
\newglossaryentry{orientationTensor}{
    type=latin,
    name={\ensuremath{\tensor{M}}},
    description={Orientation tensor},
    unit={\si{-}}
}
\newglossaryentry{orientationTensorComp}{
    type=hidden,
    name={\ensuremath{M}},
    description={Orientation tensor component},
    unit={\si{-}}
}
\newglossaryentry{fiberDirectionComp}{
    type=hidden,
    name={\ensuremath{m}},
    description={%
        Vector component of the preferred direction of a fiber reinforced
        material
    },
    unit={\si{-}}
}
\newglossaryentry{groundSubstance}{
    type=subscript,
    name={\ensuremath{g}},
    description={Ground substance}
}
\newglossaryentry{collagen}{
    type=subscript,
    name={\ensuremath{c}},
    description={Collagen fibers}
}
\newglossaryentry{wallLumenRatio}{
    type=latin,
    name={\ensuremath{\mathit{WLR}}},
    description={Wall to lumen ratio},
    unit={\si{-}}
}

\newglossaryentry{fsiInterface}{
    type=greek,
    name={\ensuremath{\gls{domainBoundary}^{\gls{fs}}}},
    description={Fluid-solid interface  = $\Lambda^f \cap \Lambda^s$},
    unit={\si{-}}
}
\newglossaryentry{fsiInterfaceFluid}{
    type=greek,
    name={\ensuremath{\gls{domainBoundary}_{fluid}^{\gls{fs}}}},
    description={%
        Fluid mesh boundary corresponding to the fluid-solid interface
    },
    unit={\si{-}}
}
\newglossaryentry{fsiInterfaceSolid}{
    type=greek,
    name={\ensuremath{\gls{domainBoundary}_{solid}^{\gls{fs}}}},
    description={%
        Solid mesh boundary corresponding to the fluid-solid interface
    },
    unit={\si{-}}
}
\newglossaryentry{fsiTolerance}{
    type=greek,
    name={\ensuremath{{\epsilon}_0}},
    description={Tolerance of the FSI coupling iterations},
    unit={\si{-}}
}
\newglossaryentry{fluidSolver}{
    type=latin,
    name={\ensuremath{\bm{\mathcal{F}}}},
    description={Symbolic representation of the fluid solver},
    unit={\si{-}}
}
\newglossaryentry{solidSolver}{
    type=latin,
    name={\ensuremath{\bm{\mathcal{S}}}},
    description={Symbolic representation of the solid solver},
    unit={\si{-}}
}
\newglossaryentry{addedMassOperator}{
    type=latin,
    name={\ensuremath{\bm{\mathcal{M}_{\mathrm{a}}}}},
    description={Symbolic representation of the added-mass operator},
    unit={\si{-}}
}
\newglossaryentry{addedMassEigenvalue}{
    type=greek,
    name={\ensuremath{\mu_\mathrm{a}}},
    description={Eigenvalue of the added-mass operator},
    unit={\si{-}}
}
\newglossaryentry{fsiTraction}{
    type=latin,
    name={\ensuremath{\cvec{y}}},
    description={%
        Solid solver input: traction exerted by the flow on the FSI interface
    },
    unit={\si{\pascal}}
}
\newglossaryentry{fsiInterfaceDispl}{
    type=latin,
    name={\ensuremath{\cvec{d}}},
    description={Position of the FSI interface},
    unit={\si{\meter}}
}
\newglossaryentry{fsDisplOutput}{
    type=latin,
    name={\ensuremath{\cvec{\tilde{d}}}},
    description={FSI interface position after a coupling iteration},
    unit={\si{\meter}}
}
\newglossaryentry{fsiResidualOperator}{
    type=latin,
    name={\ensuremath{\bm{\mathcal{R}}}},
    description={%
        Residual operator resulted from the fixed-point formulation of the FSI
        problem
    },
    unit={\si{\meter}}
}
\newglossaryentry{fsiInterfaceDoF}{
    type=latin,
    name={\ensuremath{N^{\gls{fs}}}},
    description={Number of degrees of freedom on the FSI interface},
    unit={\si{-}}
}
\newglossaryentry{fsiResidual}{
    type=latin,
    name={\ensuremath{\cvec{r}}},
    description={Discrete residual vector of the coupling iterations},
    unit={\si{\meter}}
}
\newglossaryentry{fsiReuse}{
    type=latin,
    name={\ensuremath{q}},
    description={Number of time-steps reused in the IQN-ILS algorithm},
    unit={\si{-}}
}

\newglossaryentry{peakSystole}{
    type=subscript,
    name={\ensuremath{ps}},
    description={%
        Indicates variables defined at the peak systole instant
    }
}
\newglossaryentry{lowDiastole}{
    type=subscript,
    name={\ensuremath{ld}},
    description={%
        Indicates variables defined at the low diastole instant
    }
}
\newglossaryentry{aneurysm}{
    type=subscript,
    name={\ensuremath{ia}},
    description={%
        Indicates variables defined or integrated on an aneurysm surface
    }
}
\newglossaryentry{convexHull}{
    type=subscript,
    name={\ensuremath{ch}},
    description={%
        Indicates variables defined or integrated on an aneurysm surface
        convex hull
    }
}
\newglossaryentry{ostium}{
    type=subscript,
    name={\ensuremath{o}},
    description={%
        Indicates variables defined or integrated on an aneurysm ostium
        surface
    }
}

\newglossaryentry{domeHeight}{
    type=latin,
    name={\ensuremath{h_d}},
    description={%
        Maximum height of an aneurysm's dome perpendicular to ostium
        surface
    },
    unit={\si{\meter}}
}
\newglossaryentry{maximumDomeHeight}{
    type=latin,
    name={\ensuremath{h_{max}}},
    description={%
        Maximum height of an aneurysm's dome, measured from the ostium center
    },
    unit={\si{\meter}}
}
\newglossaryentry{bulgeHeight}{
    type=latin,
    name={\ensuremath{h_b}},
    description={%
        Bulge height of an aneurysm's dome
    },
    unit={\si{\meter}}
}
\newglossaryentry{neckDiameter}{
    type=latin,
    name={\ensuremath{d_n}},
    description={Maximum diameter of an aneurysm's neck},
    unit={\si{\meter}}
}
\newglossaryentry{domeDiameter}{
    type=latin,
    name={\ensuremath{d_d}},
    description={Maximum diameter of an aneurysm's dome},
    unit={\si{\meter}}
}
\newglossaryentry{aspectRatio}{
    type=latin,
    name={\ensuremath{\mathit{AR}}},
    description={%
        Aspect ratio of an aneurysm, defined as the ratio between dome height
        and neck diameter
    },
    unit={\si{-}}
}
\newglossaryentry{meanBloodFlowRate}{
    type=latin,
    name={\ensuremath{\bar{q}_a}},
    description={Mean blood flow rate through an artery section},
    unit={\si{\milli\litre\per\second}}
}
\newglossaryentry{timeBloodFlowRate}{
    name={\ensuremath{q_a}},
    description={%
        Blood flow rate over time through an artery section
    },
    unit={\si{\milli\litre\per\second}}
}
\newglossaryentry{normBloodFlowRate}{
    name={\ensuremath{q^{*}}},
    description={%
        Normalized blood flow rate profile through the ICA section measured
        experimentally for a population.
    },
    unit={\si{-}}
}
\newglossaryentry{arteryDiameter}{
    type=latin,
    name={\ensuremath{d_{pa}}},
    description={Diameter of an aneurysm's parent artery section},
    unit={\si{\meter}}
}

\newglossaryentry{bottleneckFactor}{
    type=latin,
    name={\ensuremath{\mathit{BF}}},
    description={Bottleneck factor},
    unit={\si{-}}
}
\newglossaryentry{bulgeLocation}{
    type=latin,
    name={\ensuremath{\mathit{BL}}},
    description={Bulge location},
    unit={\si{-}}
}
\newglossaryentry{sizeRatio}{
    type=latin,
    name={\ensuremath{\mathit{SR}}},
    description={Size ratio},
    unit={\si{-}}
}
\newglossaryentry{convexityRatio}{
    type=latin,
    name={\ensuremath{\mathit{CR}}},
    description={Convexity ratio},
    unit={\si{-}}
}
\newglossaryentry{isoperimetricRatio}{
    type=latin,
    name={\ensuremath{\mathit{IPR}}},
    description={Isoperimetric ratio},
    unit={\si{-}}
}
\newglossaryentry{undulationIndex}{
    type=latin,
    name={\ensuremath{\mathit{UI}}},
    description={Undulation index},
    unit={\si{-}}
}
\newglossaryentry{ellipticityIndex}{
    type=latin,
    name={\ensuremath{\mathit{EI}}},
    description={Ellipticity index},
    unit={\si{-}}
}
\newglossaryentry{nonsphericityIndex}{
    type=latin,
    name={\ensuremath{\mathit{NSI}}},
    description={Nonsphericity index},
    unit={\si{-}}
}
\newglossaryentry{conicityParameter}{
    type=latin,
    name={\ensuremath{\mathit{CP}}},
    description={Conicity Parameter},
    unit={\si{-}}
}
\newglossaryentry{firstPrincipalCurvature}{
    type=greek,
    name={\ensuremath{\kappa_{1}}},
    description={First principal curvature of a surface},
    unit={\si{\per\meter}}
}
\newglossaryentry{secondPrincipalCurvature}{
    type=greek,
    name={\ensuremath{\kappa_{2}}},
    description={Second principal curvature of a surface},
    unit={\si{\per\meter}}
}
\newglossaryentry{meanCurvature}{
    type=latin,
    name={\ensuremath{H}},
    description={Mean curvature of a surface},
    unit={\si{\per\meter}}
}
\newglossaryentry{GaussianCurvature}{
    type=latin,
    name={\ensuremath{K}},
    description={Gaussian curvature of a surface},
    unit={\si{\per\square\meter}}
}
\newglossaryentry{areaAvgGaussianCurvature}{
    type=latin,
    name={\ensuremath{\mathit{GAA}}},
    description={Area-averaged Gaussian curvature of a surface},
    unit={\si{\per\square\milli\meter}}
}
\newglossaryentry{areaAvgMeanCurvature}{
    type=latin,
    name={\ensuremath{\mathit{MAA}}},
    description={Area-averaged mean curvature of a surface},
    unit={\si{\per\milli\meter}}
}
\newglossaryentry{l2NormGaussianCurvature}{
    type=latin,
    name={\ensuremath{\mathit{GLN}}},
    description={L2-norm of the Gaussian curvature field of a surface},
    unit={\si{-}}
}
\newglossaryentry{l2NormMeanCurvature}{
    type=latin,
    name={\ensuremath{\mathit{MLN}}},
    description={L2-norm of the mean curvature of a surface},
    unit={\si{-}}
}

\newglossaryentry{wallShearStressVector}{
    type=greek,
    name={\ensuremath{ \vec{\tau}_w }},
    description={%
        Wall shear stress vector defined on a vessel and aneurysm surface
    },
    unit={\si{\pascal}}
}
\newglossaryentry{wallShearStressMag}{
    type=hidden,
    name={\ensuremath{{\tau}_w}},
    description={%
        Wall shear stress magnitude defined on a vessel and aneurysm surface
    },
    unit={\si{\pascal}}
}
\newglossaryentry{timeAvgWallShearStress}{
    type=latin,
    name={\ensuremath{\mathit{TAWSS}}},
    description={%
        Time-average WSS field
    },
    unit={\si{\pascal\per\meter}}
}

\newglossaryentry{wallShearStressGradient}{
    type=latin,
    name={\ensuremath{\vec{G}}},
    description={%
        Wall shear stress gradient vector, defined on the surface
        $pq$-coordinate system
    },
    unit={\si{\pascal\per\meter}}
}
\newglossaryentry{wallShearStressSpatialGradient}{
    type=latin,
    name={\ensuremath{\avg{G}}},
    description={%
        Temporal-average of the wall shear stress (spatial) gradient vector
        magnitude, defined on the surface $pq$-coordinate system,
        \gls{wallShearStressGradient}
    },
    unit={\si{\pascal\per\meter}}
}
\newglossaryentry{timeAvgWallShearStressGradient}{
    type=latin,
    name={\ensuremath{\mathit{TAWSSG}}},
    description={%
        Surface gradient of the time-averaged wall shear stress gradient
        vector, defined on the surface $p$-coordinate direction (the flow
        direction)
    },
    unit={\si{\pascal\per\meter}}
}
\newglossaryentry{avgShearDirectionVector}{
    type=latin,
    name={\ensuremath{\vec{p}}},
    description={Mean flow direction adjacent to an artery's lumen},
    unit={\si{-}}
}
\newglossaryentry{transShearDirectionVector}{
    type=latin,
    name={\ensuremath{\vec{q}}},
    description={%
        Transversal direction to the mean flow direction (\vec{p})
    },
    unit={\si{-}}
}
\newglossaryentry{oscillatoryShearIndex}{
    type=latin,
    name={\ensuremath{\mathit{OSI}}},
    description={The oscillatory shear index},
    unit={\si{-}}
}
\newglossaryentry{relativeResidenceTime}{
    type=latin,
    name={\ensuremath{\mathit{RRT}}},
    description={The relative residence time indicator},
    unit={\si{-}}
}
\newglossaryentry{gradientOscillatoryNumber}{
    type=latin,
    name={\ensuremath{\mathit{GON}}},
    description={The gradient oscillatory number},
    unit={\si{-}}
}
\newglossaryentry{aneurysmFormationIndicator}{
    type=latin,
    name={\ensuremath{\mathit{AFI}}},
    description={The \textit{potential} aneurysm formation indicator},
    unit={\si{-}}
}
\newglossaryentry{cardiacPeriod}{
    type=latin,
    name={\ensuremath{T_c}},
    description={Cardiac period},
    unit={\si{\second}}
}
\newglossaryentry{lowShearArea}{
    type=latin,
    name={\ensuremath{\mathit{lsa}}},
    description={Normalized area of low WSS defined on the aneurysm surface},
    unit={\si{-}}
}
\newglossaryentry{avgParentArteryWSS}{
    type=latin,
    name={\ensuremath{\avg{\tau}_{w,pa}}},
    description={%
        Averaged parent artery WSS, defined in the anterior portion of the
        parent artery
    },
    unit={\si{\pascal}}
}
\newglossaryentry{lowWSSReference}{
    type=latin,
    name={\ensuremath{(\gls{wallShearStressMag})_{low}}},
    description={Low WSS reference},
    unit={\si{-}}
}
\newglossaryentry{wssPulsatilityIndex}{
    type=latin,
    name={\ensuremath{\mathit{WSSPI}}},
    description={Pulsatility index of the wall shear stress magnitude},
    unit={\si{-}}
}
\newglossaryentry{wssTemporalGradient}{
    type=latin,
    name={\ensuremath{\mathit{WSSTG}}},
    description={%
        Maximal temporal derivative of the wall shear stress magnitude
    },
    unit={\si{\pascal\per\second}}
}
\newglossaryentry{transWallShearStress}{
    type=latin,
    name={\ensuremath{\mathit{transWSS}}},
    description={%
        Temporal average of the wall shear stress vector along its average
        transversal direction
    },
    unit={\si{\pascal}}
}
\newglossaryentry{wssDivergence}{
    type=latin,
    name={\ensuremath{\mathrm{divWSS}}},
    description={Divergence of the normalized wall shear stress vector field},
    unit={\si[per-mode=reciprocal]{\pascal\per\milli\meter}}
}
\newglossaryentry{speedup}{
    type=latin,
    name={\ensuremath{s}},
    description={Speedup},
    unit={\si{-}}
}
\newglossaryentry{residualVector}{
    type=latin,
    name={\ensuremath{\cvec{r}}},
    description={%
        Residual vector of a linear system $\matrix{A} \cvec{x} = \cvec{b}$
    },
    unit={\si{-}}
}
\newglossaryentry{residualNorm}{
    type=hidden,
    name={\ensuremath{R_s}},
    description={%
        Scaled normalized residual of a linear system in an iterative solver
    },
    unit={\si{-}}
}
\newglossaryentry{scaleFactor}{
    type=latin,
    name={\ensuremath{F_s}},
    description={Scale factor used for scaling the residual norm},
    unit={\si{-}}
}
\newglossaryentry{identityMatrix}{
    type=hidden,
    name={\ensuremath{\matrix{I}}},
    description={Identity matrix of $\realnumbers^{N \times N}$},
    unit={\si{-}}
}

\newglossaryentry{turbIntegralLength}{
    type=latin,
    name={\ensuremath{\mathcal{L}}},
    description={Length integral scale of turbulent flow},
    unit={\si{\meter}}
}
\newglossaryentry{turbIntegralTime}{
    type=latin,
    name={\ensuremath{\mathcal{T}}},
    description={Time integral scale of turbulent flow},
    unit={\si{\second}}
}
\newglossaryentry{turbIntegralVelocity}{
    type=latin,
    name={\ensuremath{\mathcal{V}}},
    description={Velocity integral scale of turbulent flow},
    unit={\si{\meter\per\second}}
}
\newglossaryentry{KolmogorovLength}{
    type=latin,
    name={\ensuremath{\eta}},
    description={Kolmogorov length scale of turbulent flow},
    unit={\si{\meter}}
}
\newglossaryentry{KolmogorovTime}{
    type=latin,
    name={\ensuremath{\tau}},
    description={KOlmogorov time scale of turbulent flow},
    unit={\si{\second}}
}
\newglossaryentry{KolmogorovVelocity}{
    type=latin,
    name={\ensuremath{\upsilon}},
    description={Kolmogorov velocity scale of turbulent flow},
    unit={\si{\meter\per\second}}
}
\newglossaryentry{fluctuationVelocity}{
    type=latin,
    name={\ensuremath{\vec{u^{'}}}},
    description={Vetor de velocidade flutuante},
    unit={\si{\meter\per\second}}
}
\newglossaryentry{meanVelocity}{
	type=latin,	
    name={\ensuremath{\overline{\gls{velocityComp}}}},				
	description={Vetor de velocidade média},
	unit={\si{\meter\per\second}}
}
\newglossaryentry{yPlus}{
    type=latin,
    name={\ensuremath{y^+}},
    description={Dimensionless wall distance},
    unit={-}
}
\newglossaryentry{turbulenceIntensity}{
    type=latin,
    name={\ensuremath{I_t}},
    description={Turbulence intensity for inlet boundaries},
    unit={\si{-}}
}

\newglossaryentry{firstCellSize}{
    type=latin,
    name={\ensuremath{h}},
    description={The distance nearest to the wall of the first cell},
    unit={\si{\meter}}
}

\newglossaryentry{boundaryLayerThickness}{
    type=latin,
    name={\ensuremath{\delta}},
    description={Boundary-layer thickness},
    unit={\si{\meter}}
}
\newglossaryentry{velocityPlus}{
    type=latin,
    name={\ensuremath{u^{+}}},
    description={Dimensionless velocity in the turbulent boundary layer},
    unit={\si{\meter\per\second}}
}
\newglossaryentry{frictionVelocity}{
    type=latin,
    name={\ensuremath{u_{\tau}}},
    description={Friction velocity},
    unit={\si{\meter\per\second}}
}
\newglossaryentry{turbKineticEnergy}{
    type=latin,
    name={\ensuremath{k}},
    description={Energia cinética turbulenta por unidade de massa},
    unit={\si{\square\meter\per\square\second}}
}

\newglossaryentry{turbDissipationRate}{
    type=greek,
    name={\ensuremath{\varepsilon}},
    description={Taxa de dissipação da energia cinética turbulenta},
    unit={\si{\square\meter\per\cubic\second}}
}

\newglossaryentry{turbDynamicViscosity}{
    type=greek,
    name={\ensuremath{\mu_{\text{t}}}},
    description={Viscosidade din\^amica turbulenta},
    unit={\si{\pascal\second}}
}
\newglossaryentry{turbViscosity}{
    type=greek,
    name={\ensuremath{\nu_{\text{t}}}},
    description={Viscosidade cinemática turbulenta},
    unit={\si{\square\meter\per\second}}
}
\newglossaryentry{eddyViscosityParameter}{
    type=greek,
    name={\ensuremath{\tilde{\nu}}},
    description={%
        Kinematic eddy viscosity parameter of the Spalart-Allmaras models%
    },
    unit={\si{\square\meter\per\second}}
}
\newglossaryentry{wallDampingFunction}{
    type=greek,
    name={\ensuremath{f_{\nu 1}}},
    description={%
        Wall-damping function of the Spalart-Allmaras model%
    },
    unit={\si{-}}
}

\newglossaryentry{effTurbViscosity}{
    type=greek,
    name={\ensuremath{\nu_{\text{ef}}}},
    description={Viscosidade cinemática efetiva},
    unit={\si{\square\meter\per\second}}
}

\newglossaryentry{effkDiffusivity}{
    type=latin,
    name={\ensuremath{d_{\gls{turbKineticEnergy},\text{ef}}}},
    description={Difusividade efetiva de $k$},
    unit={\si{\square\meter\per\second}}
}

\newglossaryentry{effEpsilonDiffusivity}{
    type=latin,
    name={\ensuremath{d_{\gls{turbDissipationRate},\text{ef}}}},
    description={Difusividade efetiva de $\epsilon$},
    unit={\si{\square\meter\per\second}}
}

\newglossaryentry{effOmegaDiffusivity}{
    type=latin,
    name={\ensuremath{d_{\omega,\text{ef}}}},
    description={Difusividade efetiva de $\omega$},
    unit={\si{\square\meter\per\second}}
}

\newglossaryentry{effPhiDiffusivity}{
    type=latin,
    name={\ensuremath{d_{\phi,\text{ef}}}},
    description={Difusividade efetiva de $\phi$},
    unit={\si{\square\meter\per\second}}
}

\newglossaryentry{turbKineticEnergyProduction}{
    type=latin,
    name={\ensuremath{P_{\gls{turbKineticEnergy}}}},
    description={Produção de energia cinética turbulenta},
    unit={\si{\square\meter\per\cubic\second}}
}

\newglossaryentry{limitedTurbProduction}{
    type=latin,
    name={\ensuremath{P_{k}^{'}}},
    description={Produção de energia cinética turbulenta limitado},
    unit={\si{\square\meter\per\cubic\second}}
}

\newglossaryentry{limitedOmegaTurbProduction}{
    type=latin,
    name={\ensuremath{P_{\omega}^{'}}},
    description={Produção de energia cinética turbulenta limitado},
    unit={\si{\square\meter\per\cubic\second}}
}

\newglossaryentry{specificDissipationRate}{
    type=greek,
    name={\ensuremath{\omega}},
    description={Turbulence frequency},
    unit={\si{\per\second}}
}

\newglossaryentry{F1}{
    type=latin,
    name={\ensuremath{F_{1}}},
    description={Função do modelo k-$\omega$ SST},
    unit={-}
}

\newglossaryentry{F2}{
    type=latin,
    name={\ensuremath{F_{2}}},
    description={Função do modelo k-$\omega$ SST},
    unit={-}
}

\newglossaryentry{nearestDistWall}{
    type=latin,
    name={\ensuremath{y_{\perp}}},
    description={Distância mais próxima da parede},
    unit={\si{\meter}}
}

\newglossaryentry{cdkw}{
    type=latin,
    name={\ensuremath{cd_{k\omega}}},
    description={Parte positiva do termo cruzado de difusão da equação de transporte de $\omega$},
    unit={\si{1\per\second^{4}}}
}

\newglossaryentry{psi}{
    type=latin,
    name={\ensuremath{\psi}},
    description={Coeficientes do modelo de turbulência k-$omega$ SST},
    unit={-}
}

\newglossaryentry{cmu}{
    type=latin,
    name={\ensuremath{C_{\mu}}},
    description={Constante do modelo de turbulência},
    unit={-}
}

\newglossaryentry{kPrandtlNumber}{
    type=greek,
    name={\ensuremath{\sigma_{k}}},
    description={Número de Prandtl turbulento para $k$},
    unit={-}
}

\newglossaryentry{epsilonPrandtlNumber}{
    type=greek,
    name={\ensuremath{\sigma_{\epsilon}}},
    description={Número de Prandtl turbulento para $\epsilon$},
    unit={-}
}
\newglossaryentry{turbFrequencyPrandtlNumber}{
    type=greek,
    name={\ensuremath{\sigma_{\omega}}},
    description={Turbulence frequency Prandtl number},
    unit={-}
}

\newglossaryentry{alphaPhi}{
    type=greek,
    name={\ensuremath{\alpha_{\phi}}},
    description={Inverso do número de Prandtl turbulento},
    unit={-}
}

\newglossaryentry{a1}{
    type=latin,
    name={\ensuremath{a_{1}}},
    description={Constante do modelo de turbulência},
    unit={-}
}

\newglossaryentry{b1}{
    type=latin,
    name={\ensuremath{b_{1}}},
    description={Constante do modelo de turbulência},
    unit={-}
}

\newglossaryentry{c1}{
    type=latin,
    name={\ensuremath{c_{1}}},
    description={Constante do modelo de turbulência},
    unit={-}
}

\newglossaryentry{C1epsilon}{
    type=latin,
    name={\ensuremath{C_{1\gls{turbDissipationRate}}}},
    description={Constante do modelo de turbulência},
    unit={-}
}

\newglossaryentry{C2epsilon}{
    type=latin,
    name={\ensuremath{C_{2\gls{turbDissipationRate}}}},
    description={Constante do modelo de turbulência},
    unit={-}
}

\newglossaryentry{beta}{
    type=greek,
    name={\ensuremath{\beta}},
    description={Constante do modelo de turbulência},
    unit={-}
}

\newglossaryentry{betaStar}{
    type=greek,
    name={\ensuremath{\beta^{*}}},
    description={Constante do modelo de turbulência},
    unit={-}
}

\newglossaryentry{gamma}{
    type=greek,
    name={\ensuremath{\gamma}},
    description={Constante do modelo de turbulência},
    unit={-}
}

\newglossaryentry{sigmaOmega1}{
    type=greek,
    name={\ensuremath{\sigma_{\omega 1}}},
    description={Constante do modelo de turbulência},
    unit={-}
}

\newglossaryentry{sigmaOmega2}{
    type=greek,
    name={\ensuremath{\sigma_{\omega 2}}},
    description={Constante do modelo de turbulência},
    unit={-}
}

\newglossaryentry{alphak1}{
    type=greek,
    name={\ensuremath{\alpha_{k 1}}},
    description={Constante do modelo de turbulência},
    unit={-}
}

\newglossaryentry{alphak2}{
    type=greek,
    name={\ensuremath{\alpha_{k 2}}},
    description={Constante do modelo de turbulência},
    unit={-}
}

\newglossaryentry{alphaOmega1}{
    type=greek,
    name={\ensuremath{\alpha_{\omega 1}}},
    description={Constante do modelo de turbulência},
    unit={-}
}

\newglossaryentry{alphaOmega2}{
    type=greek,
    name={\ensuremath{\alpha_{\omega 2}}},
    description={Constante do modelo de turbulência},
    unit={-}
}

\newglossaryentry{beta1}{
    type=greek,
    name={\ensuremath{\beta_{1}}},
    description={Constante do modelo de turbulência},
    unit={-}
}

\newglossaryentry{beta2}{
    type=greek,
    name={\ensuremath{\beta_{2}}},
    description={Constante do modelo de turbulência},
    unit={-}
}

\newglossaryentry{gamma1}{
    type=greek,
    name={\ensuremath{\gamma_{1}}},
    description={Constante do modelo de turbulência},
    unit={-}
}

\newglossaryentry{gamma2}{
    type=greek,
    name={\ensuremath{\gamma_{2}}},
    description={Constante do modelo de turbulência},
    unit={-}
}
\newglossaryentry{ReynoldsTensor}{
    type=greek,
    name={\ensuremath{\tensor{\tau^{R}}}},
    description={Tensor das tensões de Reynolds},
    unit={\si{\per\second}}
}
\newglossaryentry{ReynoldsTensorComp}{
    type=hidden,
    name={\ensuremath{\tau^{R}}},
    description={Tensor das tensões de Reynolds},
    unit={\si{\per\second}}
}

\newglossaryentry{meanStrainRateTensor}{
    type=latin,
    name={\ensuremath{\overline{\gls{strainRateTensor}}}},
    description={Tensor taxa de deformação média},
    unit={\si{\per\second}}
}

\newglossaryentry{mesh}{
    type=hidden,
    name={\ensuremath{M}},
    description={%
        Símbolo genérico representando uma malha computacional
    },
    unit={\si{-}}
}
\newglossaryentry{coarse}{
    type=subscript,
    name={\ensuremath{c}},
    description={%
        Indica propriedades da malha mais grosseira (\textit{coarse})
    },
    unit={\si{-}}
}
\newglossaryentry{intermediate}{
    type=subscript,
    name={\ensuremath{i}},
    description={%
        Indica propriedades da malha intermediária
    },
    unit={\si{-}}
}
\newglossaryentry{fine}{
    type=subscript,
    name={\ensuremath{f}},
    description={%
        Indica propriedades da malha mais refinada (\textit{finer})
    },
    unit={\si{-}}
}
\newglossaryentry{meshCellsNumber}{
    type=hidden,
    name={\ensuremath{N_M}},
    description={%
        Number of cells in a computational mesh
    },
    unit={\si{-}}
}
\newglossaryentry{meshSurfaceCellDensity}{
    type=hidden,
    name={\ensuremath{\rho_{S}}},
    description={%
        Cells surface density in a computational mesh
    },
    unit={\si{cells\per\milli\square\meter}}
}
\newglossaryentry{meshVolumeCellDensity}{
    type=hidden,
    name={\ensuremath{\rho_{V}}},
    description={%
        Cells volumetric density in a computational mesh
    },
    unit={\si{cells\per\milli\cubic\meter}}
}
\newglossaryentry{maxEdgeSize}{
    type=hidden,
    name={\ensuremath{l_e}},
    description={%
        Maximum edge length size in a computational mesh
    },
    unit={\si{\milli\meter}}
}
\newglossaryentry{meshSize}{
    type=hidden,
    name={\ensuremath{h}},
    description={%
        Edge length size of a generic structured computational mesh
    },
    unit={\si{\milli\meter}}
}
\newglossaryentry{meshRefinementRatio}{
    type=hidden,
    name={\ensuremath{r}},
    description={%
        Ratio between element size of systematically refined meshes
    },
    unit={\si{-}}
}
\newglossaryentry{fvmOrder}{
    type=hidden,
    name={\ensuremath{p}},
    description={%
        Formal order of accuracy of the Finite Volume Method
    },
    unit={\si{-}}
}
\newglossaryentry{realFvmOrder}{
    type=hidden,
    name={\ensuremath{\tilde{p}}},
    description={%
        Real or practical order of accuracy of the Finite Volume Method
    },
    unit={\si{-}}
}
\newglossaryentry{solutionDifference}{
    type=latin,
    name={\ensuremath{\epsilon}},
    description={%
        Diferença (absoluta ou relativa) entre as soluções de duas malhas
        sistematicamente refinadas
    },
    unit={\si{-}}
}
\newglossaryentry{discretizationError}{
    type=latin,
    name={\ensuremath{E^{\mathrm{R}}}},
    description={%
        Discretization error of a particular mesh computed using Richardson's
        extrapolation
    },
    unit={\si{-}}
}
\newglossaryentry{truncationError}{
    type=latin,
    name={\ensuremath{e^{\mathrm{t}}}},
    description={%
        Truncation error of a discretization scheme.
    },
    unit={\si{-}}
}
\newglossaryentry{gridConvergenceIndex}{
    type=latin,
    name={\ensuremath{\mathit{GCI}}},
    description={%
        Grid convergence index
    },
    unit={\si{-}}
}
\newglossaryentry{gciSafetyFactor}{
    type=latin,
    name={\ensuremath{F_s}},
    description={%
        Safety factor defined with the grid convergence index
    },
    unit={\si{-}}
}

\newacronym[category=standard,type=hidden]{1d}{1D}{one-dimensional}
\newacronym[category=standard,type=hidden]{2d}{2D}{two-dimensional}
\newacronym[category=standard,type=hidden]{3d}{3D}{three-dimensional}
\newacronym[category=standard,type=hidden]{4d}{4D}{four-dimensional}

\newacronym[category=standard,type=hidden]{vmtk}{VMTK}{Vascular Modeling Toolkit}
\newacronym[category=standard,type=hidden]{vtk}{VTK}{Visualization Toolkit}

\newacronym{cfd}{CFD}{Computational Fluid Dynamics}
\newacronym{csd}{CSD}{Computational Solid Dynamics}
\newacronym{cae}{CAE}{Computer-Aided Engineering}
\newacronym{fea}{FEA}{Finite Element Analysis}
\newacronym{cem}{CEM}{Computational Electromagnetic}
\newacronym{mbd}{MBD}{Multi-body Dynamics}
\newacronym{fsi}{FSI}{Fluid-Solid Interaction}

\newacronym{pde}{PDE}{partial differential equations}
\newacronym{fvm}{FVM}{Finite Volume Method}
\newacronym{fdm}{FDM}{Finite Difference Method}
\newacronym{fem}{FEM}{Finite Element Method}
\newacronym{ale}{ALE}{arbitrary Lagrangian-Eulerian}
\newacronym{iqn-ils}{IQN-ILS}{Interface Quasi-Newton-Implicit Jacobian Least-Squares}
\newacronym[category=never-short,type=hidden]{bc}{BC}{boundary condition}

\newacronym[category=only-short,type=hidden]{simple}{SIMPLE}{Semi-Implicit Method for Pressure Linked Equations}
\newacronym[category=only-short,type=hidden]{piso}{PISO}{Pressure Implicit with Splitting Operators}

\newacronym[category=never-short,type=hidden]{am}{AM}{added-mass}
\newacronym[category=never-short,type=hidden]{dnf}{DNF}{Dirichlet-Neumann formulation}
\newacronym[category=never-short,type=hidden]{dof}{DOF}{degrees of freedom}
\newacronym[category=never-short,type=hidden]{pae}{PAE}{peri-aneurysmal environment}

\newacronym{ia}{CA}{cerebral aneurysm}
\newacronym{aaa}{AAA}{abdominal aortic aneurysm}
\newacronym{voi}{VOI}{volume of interest}
\newacronym{dsa}{DSA}{digital subtraction angiography}
\newacronym{cta}{CTA}{computational tomography angiography}
\newacronym{wlr}{WLR}{wall-to-lumen ratio}
\newacronym{sah}{SAH}{subarachnoid hemorrhage}

\newacronym{wss}{WSS}{wall shear stress}
\newacronym{tawss}{TAWSS}{time-averaged wall shear stress}
\newacronym{wsstg}{WSSTG}{maximum wall shear stress temporal gradient}
\newacronym{wsssg}{WSSSG}{wall shear stress surface gradient}
\newacronym{transwss}{transWSS}{transversal wall shear stress}
\newacronym{osi}{OSI}{oscillatory shear index}
\newacronym{afi}{AFI}{aneurysm formation indicator}
\newacronym{lsa}{LSA}{low shear area}

\newacronym[category=never-short,type=hidden]{pswss}{PSWSS}{peak-systole WSS}
\newacronym[category=never-short,type=hidden]{gon}{GON}{gradient oscillatory number}
\newacronym[category=never-short,type=hidden]{rrt}{RRT}{relative residence time}
\newacronym[category=never-short,type=hidden]{wsspi}{WSSPI}{wall shear stress pulsatility index}
\newacronym{tawssg}{TAWSSG}{time-averaged WSS gradient}
\newacronym[category=never-short,type=hidden]{pica}{PICA}{posterior inferior cerebellar artery}
\newacronym[category=never-short,type=hidden]{pca}{PCA}{posterior cerebral artery}

\newacronym[category=never-short,type=hidden]{mca}{MCA}{middle cerebral artery}
\newacronym[category=never-short,type=hidden]{ica}{ICA}{internal carotid artery}
\newacronym[category=never-short,type=hidden]{aca}{ACA}{anterior cerebral artery}
\newacronym[category=never-short,type=hidden]{acoa}{ACoA}{anterior communicating artery}
\newacronym[category=never-short,type=hidden]{ba}{BA}{basilar artery}
\newacronym[category=never-short,type=hidden]{va}{VA}{vertebral arteries}
\newacronym[category=never-short,type=hidden]{oa}{OA}{ophthalmic artery}

\newacronym[category=standard,type=hidden]{fapesp}{FAPESP}{São Paulo Research Foundation}
\newacronym[category=standard,type=hidden]{unesp}{UNESP}{São Paulo State University}

\newacronym[category=standard,type=hidden]{mri}{MRI}{magnetic resonance imaging}
\newacronym[category=never-short,type=hidden]{1wfsi}{1WFSI}{one-way fluid-solid interaction}

\newacronym[category=standard,type=hidden]{uiats}{UIATS}{unruptured \gls{ia} treatment score}
\newacronym[category=standard,type=hidden]{rrs}{RRS}{rupture resamblance score}
\newacronym[category=never-short,type=hidden]{uwm}{UWM}{uniform-wall model}
\newacronym[category=never-short,type=hidden]{awm}{AWM}{abnormal-wall model}
\newacronym{sd}{SD}{standard deviation}

\newglossaryentry{vasculature}{
    type=subscript,
    name={\ensuremath{v}},
    description={%
        Indicates variables defined or integrated on the vasculature surface
    }
}
\newglossaryentry{parentArtery}{
    type=subscript,
    name={\ensuremath{pa}},
    description={%
        Indicates variables defined or integrated on the parent artery surface
        of an aneurysm
    }
}
\newglossaryentry{branch}{
    type=subscript,
    name={\ensuremath{b}},
    description={%
        Indicates variables defined on the branches of a vasculature
    }
}
\newglossaryentry{populationAvgDiameter}{
    type=latin,
    name={\ensuremath{d_{avg}}},
    description={%
        Populational-average diameter of the ICA section
    },
    unit={\si{\milli\meter}}
}
\newglossaryentry{aneurysmPulsatility}{
    type=greek,
    name={\ensuremath{\delta_{v}}},
    description={Pulsatility of an intracranial aneurysm},
    unit={\si{-}}
}
\newglossaryentry{distanceToNeckLine}{
    type=latin,
    name={\ensuremath{g_n}},
    description={%
        Local geodesic distance to the aneurysm neck contour
    },
    unit={\si{\milli\meter}}
}
\newglossaryentry{distanceToNeckLineMax}{
    type=latin,
    name={\ensuremath{g_{n,max}}},
    description={%
        Maximum local geodesic distance to the aneurysm neck contour
    },
    unit={\si{\milli\meter}}
}
\newglossaryentry{lumenDiameter}{
    type=latin,
    name={\ensuremath{d_l}},
    description={%
        Vasculature lumen diameter at a specified section
    },
    unit={\si{\milli\meter}}
}
\newglossaryentry{intracranialPressure}{
    type=latin,
    name={\ensuremath{\gls{pressure}_{o}}},
    description={Intracranial pressure},
    unit={\si{\pascal}}
}
\newglossaryentry{patientBloodFlowRate}{
    name={\ensuremath{q_p}},
    description={%
        Blood flow rate over time through an artery section of a particular
        patient
    },
    unit={\si{\milli\litre\per\second}}
}
\newglossaryentry{ellipticPatch}{
    type=subscript,
    name={\ensuremath{E}},
    description={%
        Elliptic patches of a surface
    }
}
\newglossaryentry{hyperbolicPatch}{
    type=subscript,
    name={\ensuremath{H}},
    description={%
        Hyperbolic patches of a surface
    }
}
\newglossaryentry{rupturedGroup}{
    type=latin,
    name={\ensuremath{\mathcal{R}}},
    description={%
        The group of ruptured aneurysms
    },
    unit={\si{-}}
}
\newglossaryentry{unrupturedGroup}{
    type=latin,
    name={\ensuremath{\mathcal{U}}},
    description={%
        The group of unruptured aneurysms
    },
    unit={\si{-}}
}
\newglossaryentry{pValue}{
    type=hidden,
    name={\ensuremath{p}},
    description={P-value of statistical tests},
    unit={\si{-}}
}
\newglossaryentry{significanceLevel}{
    type=hidden,
    name={\ensuremath{\alpha}},
    description={Significance level for statistical tests},
    unit={\si{-}}
}
\newglossaryentry{standardDeviation}{
    type=hidden,
    name={\ensuremath{\mathit{SD}}},
    description={Standard-deviation of a sample's distribution},
    unit={\si{-}}
}
\newglossaryentry{SpearmanCoeff}{
    type=greek,
    name={\ensuremath{\rho_{\mathrm{S}}}},
    description={Spearman's correlation coefficient},
    unit={\si{-}}
}
\newglossaryentry{PearsonCoeff}{
    type=greek,
    name={\ensuremath{\rho_{\mathrm{P}}}},
    description={Pearson's correlation coefficient},
    unit={\si{-}}
}
\newglossaryentry{lambda2VortexId}{
    type=greek,
    name={\ensuremath{\lambda_{2}}},
    description={Lambda2 parameter of vortex identification},
    unit={\si[per-mode=reciprocal]{\per\second\squared}}
}

    \makeglossaries

    \usepackage{chngcntr}

    \usepackage[capitalise]{cleveref}

\begin{document}

    \begin{frontmatter}

    % Add title here as required by template
    \title{\ManuscriptTitle}

    \author[feb-short]{M. P. A. Dallavanzi}
    \ead{marcella.avanzi@unesp.br}

    \author[feis-short]{J.L. Gasche}
    \ead{jose.gasche@unesp.br}

    \author[feb-long]{I. L. Oliveira\corref{cor}}
    \ead{iago.oliveira@unesp.br}

    %% Author affiliation
    %\affiliation{%
    %    % Department and Organization
    %    organization={%
    %        Department of Mechanical Engineering, School of Engineering, São
    %        Paulo State University (UNESP)
    %    },%
    %    addressline={},
    %    city={},
    %    postcode={},
    %    state={},
    %    country={}%
    %}

    \cortext[cor]{Corresponding author}

    \address[feb-long]{%
        São Paulo State University (UNESP), School of Engineering, Bauru, %
        Department of Mechanical Engineering. %
        Av. Engenheiro Luiz Edmundo Carrijo Coube, 14-01, 17033-360, Bauru, SP,
        Brazil %
    }

    \address[feb-short]{%
        São Paulo State University (UNESP), School of Engineering, Bauru, %
        Department of Mechanical Engineering. %
    }

    \address[feis-short]{%
        São Paulo State University (UNESP), School of Engineering, Ilha
        Solteira, Mechanical Engineering Department
    }

    \begin{abstract}
        The relationship between vascular morphology and hemodynamics is
        important to understanding the natural history of \glspl{ia}. While
        global geometric indices have been widely studied, the local
        interaction between luminal curvature and \gls{wss} remains poorly
        characterized. This study analyzed a large, heterogeneous cohort of
        patient-specific \glspl{ia} to investigate how local surface morphology
        is related to hemodynamic environments. This was performed via
        \glsxtrlong{cfd} flow simulations of a set of 76 patient-specific
        \glspl{ia} geometries using the OpenFOAM\R library. Geometry and
        patient-specific pulsatile inflow conditions were modeled based on the
        patient arterial diameter and age. Blood was assumed as a Newtonian
        incompressible fluid flowing in a laminar regime. We utilized a
        geometric framework to classify the aneurysm lumen into spherical-like
        (elliptic) and saddle-like (hyperbolic) patches based on Gaussian
        curvature. Our results demonstrate a robust, statistically significant
        correlation between these curvature types and hemodynamic metrics,
        regardless of rupture status or aneurysm type. Specifically,
        saddle-like patches, predominantly found at the aneurysm neck, are
        associated with high \glsxtrlong{tawss}, low \glsxtrlong{osi}, and
        intense near-wall vortical activity as identified by the
        \gls{lambda2VortexId}-criterion. In contrast, spherical-like patches,
        dominant at the dome, correspond to regions of flow impingement
        characterized by lower \glsxtrlong{tawss} and elevated
        \glsxtrlong{osi}. These findings suggest that wall curvature is a
        primary determinant of local hemodynamics, overriding variations in
        global flow patterns. By bridging the gap between wall local morphology
        and pathological wall markers, this work suggests curvature-based
        mapping can serve as a powerful tool for identifying vulnerable regions
        susceptible to thinning and rupture. This objective geometric
        assessment offers valuable insights for risk stratification and the
        precision planning of endovascular interventions.
    \end{abstract}

    %% Graphical abstract
    %\begin{graphicalabstract}
    %    %\includegraphics{grabs}
    %\end{graphicalabstract}

    % Research highlights
    \begin{highlights}
        \item Local wall curvature is a robust, primary determinant of
            hemodynamics in intracranial aneurysms, independent of global flow
            patterns or rupture status.

        \item Saddle-like (hyperbolic) geometries preferentially exhibit high
            \gls{tawss} and low \gls{osi} driven by intense near-wall vortical
            activity.

        \item Spherical-like (elliptic) geometries are dominated by flow
            impingement and stagnation, leading to elevated oscillatory shear
            and lower shear stress.
    \end{highlights}


\end{frontmatter}

    \glsresetall

    \section{Introduction} \label{sec:introduction}

    The rupture of a \gls{ia} --- pathological dilatations of the arterial
    tree supplying the brain --- remains a severe neurovascular problem
    associated with high risks of mortality and morbidity
    \citep{Vlak2013,Saqr2019}. While therapeutic interventions exist, their
    significant procedural risks force neurosurgeons into a complex
    decision-making process between intervention and conservative longitudinal
    monitoring. Currently, clinical management relies on patient-specific
    factors --- such as age, sex, family history, comorbidities, and smoking
    status --- alongside fundamental geometric metrics --- aneurysm sac
    diameter and height --- and anatomical location \citep{Tjoumakaris2024}.
    Qualitative morphological features, including \enquote{blebs} or lobular
    regions, are also considered due to their strong association with increased
    rupture risk \citep{Leemans2019,Juchler2020,Levitt2021}. However, the final
    clinical assessment often remains inconclusive. Elaborated scoring systems
    have been developed to provide objective guidance
    \citep{Raghavan2005,Dhar2008,Greving2014}. For example, \citet{Etminan2015}
    proposed the \enquote{\glsxtrlong{uiats}} score based on 29 clinical,
    morphologic and geometric factors. However, an evaluation by
    \citet{RajabzadehOghaz2020a} demonstrated that \SI{43}{\percent} of cases
    were classified as non-definitive. This high rate of ambiguity suggests
    that established scoring criteria may be insufficient for a final clinical
    decision and underscores the necessity to incorporate additional factors.

    Extensive research suggests that the initiation, development, and eventual
    rupture of \glspl{ia} are primarily driven by hemodynamic forces acting on
    the vessel lumen \citep{Frosen2014,Fukazawa2015,Cebral2017}, implying that
    hemodynamic data should be incorporated into clinical scores. However, the
    specific mechanisms governing aneurysm progression remain debated
    \citep{Levitt2021}. To investigate these, \gls{cfd} has become a standard
    tool for simulating blood flow within both idealized and patient-specific
    vascular networks, elucidating variables linked to aneurysm evolution
    \citep{Geers2017,Lauric2014}. Recent investigations have shifted toward
    granular descriptions of intra-aneurysmal flow
    \citep{Ardakani2019,Mazzi2025} and localized phenomena like bleb and
    thrombus formations, which are intimately connected with specific near-wall
    flow environments \citep{SalimiAshkezari2020,Karnam2024}. Nevertheless,
    implementing \gls{cfd} in routine practice is restricted by high
    computational costs, the time-to-result for high-fidelity simulations, and
    the need for specialized expertise. Consequently, proxies for
    intra-aneurysmal flow would be valuable for faster, more accessible
    clinical assessments.

    Morphological changes, such as blebs and thrombi, alter the luminal
    curvature --- an important mathematical determinant of wall stress
    \citep{Humphrey2000,Ma2007}. Recent research has further highlighted the
    role of curvature in shaping wall stress and local hemodynamics.
    \citet{Oliveira2023} demonstrated that specific curvature types correlate
    with abnormal flow conditions known to drive aneurysm growth, rupture, and
    tissue degradation \citep{Meng2014,Frosen2019}. The present study expands
    the current understanding by investigating the the relationship between
    wall curvature and hemodynamics in a robust, larger cohort. To this end, we
    conducted \gls{cfd} simulations across 76 patient-specific \glspl{ia} under
    realistic boundary conditions. Our analysis elucidates the correlations
    between curvature metrics and hemodynamic variables while pinpointing the
    specific flow structures within the aneurysm sac that drive these
    pathological signatures. If luminal curvature dictates local hemodynamics,
    it represents a promising candidate for clinical scoring systems, offering
    an objective metric for features currently assessed only qualitatively.

    \section{Numerical Methodology} \label{sec:numericalMethodology}

\subsection{Geometries of Patient-specific Aneurysms}
\label{sec:geometries}

    Patient-specific \gls{ia} geometries were obtained from the \emph{Aneurisk}
    repository \citep{aneurisk}, which provides 100 cases under a
    \enquote{Creative Commons} license. From this dataset, 76 cases from the
    anterior circulation (including \gls{ica}, \gls{mca}, and \gls{aca}) were
    selected. The exclusion criteria were: (i) arterial trees with multiple
    aneurysms; (ii) models where outlets were positioned too close to the
    aneurysm sac to ensure accurate boundary condition application; and (iii)
    \gls{ica} lengths insufficient for consistent clipping.

    The geometries in the Aneurisk repository often include extensive arterial
    segments beyond the region of interest. To avoid operator bias during inlet
    selection, we employed an automated clipping procedure. Surfaces were
    clipped at the fourth \gls{ica} bend, identified using the curvature and
    torsion-based classification by \citet{Piccinelli2011}. This standardized
    approach ensures that flow inlets are consistently positioned across all
    cases, following the methodological recommendations of
    \citet{Valen-Sendstad2015}.

\subsection{%
    Physical Modeling of the Blood Flow and Boundary Conditions%
}   \label{sec:physicalModelFlow}

\begin{figure}[!htb]
        \centering
        \caption{%
            (a) Schematic example of the domain representing the blood flow
            through a typical \gls{ia}. (b) Population-averaged inflow rate
            waveforms used in the simulations for young adults, after data
            taken from \citet{Ford2005}, and older adults, after data taken
            from \citet{Hoi2010} (time axis normalized by the cardiac period,
            \gls{cardiacPeriod}, of each profile).
        }

        \includegraphics[width=\textwidth]{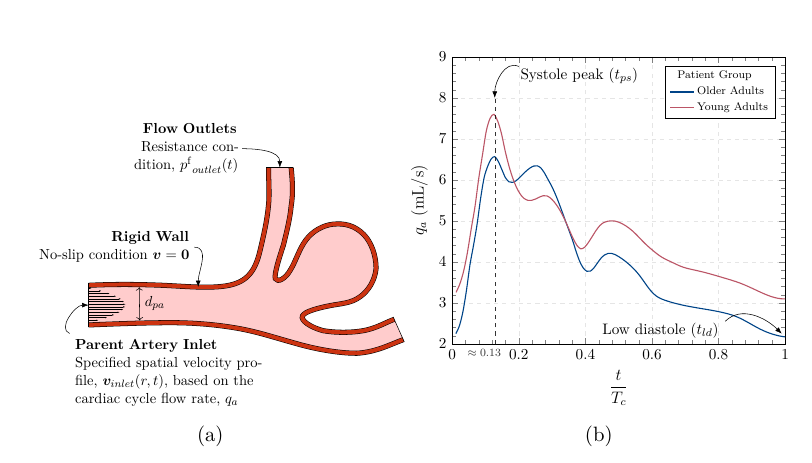}

        \label{fig:boundaryConditions}
    \end{figure}

    Blood flow was computed assuming rigid walls
    (\cref{fig:boundaryConditions}a) and an incompressible, Newtonian fluid
    model under isothermal conditions. The continuity equation:
    \begin{equation} \label{eq:continuityEquation}
        \div{
            \gls{velocity}
        } = 0\,,
    \end{equation}
    and the Navier-Stokes equations for constant viscosity:
    \begin{equation} \label{eq:momentumEquation}
        \ddt{
            \left(
                \gls{density}
                \gls{velocity}
            \right)
        }
        +
        \div{
            \left(
                \gls{density}
                \gls{velocity}
                \gls{velocity}
            \right)
        }
        =
        -\grad{\gls{pressure}}
        +
        \gls{dynamicViscosity}
        \laplacian{
            \gls{velocity}
        }\,,
    \end{equation}
    governing the fluid flow were numerically solved for the velocity,
    \gls{velocity}, and pressure, \gls{pressure}, fields. Blood was
    characterized by a density (\gls{density}) of
    \SI{1056}{\kilogram\per\cubic\meter} and dynamic viscosity
    (\gls{dynamicViscosity}) of \SI{3.5e-3}{\pascal\second}
    \citep{Shibeshi2005,Isaksen2008}.

    As the flow through the real intracranial arteries is already developed, at
    the flow inlets, we imposed a spatially-varying pulsatile velocity profile
    according to the fully-developed laminar flow in a straight pipe:
    \begin{equation} \label{eq:parabolicInletCondition}
        \gls{velocity}_{inlet}
        \functionOf{
            \gls{radial},
            \gls{time}
        }
        =
        2\frac{
            \gls{patientBloodFlowRate}
            \functionOf{\gls{time}}
        }{
            \gls{surfaceArea}_{inlet}
        }
        \left[
            1
            -
            \frac{
                4\gls{radial}^2
            }{
                \gls{arteryDiameter}^2
            }
        \right]\,,
    \end{equation}
    where $\gls{surfaceArea}_{inlet}$ is the cross-sectional area of the inlet
    artery, \gls{arteryDiameter} is its inner diameter, and \gls{radial} is the
    radial coordinate of the circular inlet section. To avoid artificial
    recirculation at the model inlet, as they are not circular, an artificial
    circular-section extension, with a length equal to twice the diameter
    \gls{arteryDiameter}, was added to the artery inlet to impose this inlet
    flow condition. The blood flow rate,
    \gls{patientBloodFlowRate}\functionOf{\gls{time}}, was determined based on
    the geometry inlet diameter and the patient age, available in the
    \emph{Aneurisk} repository data, to utilize inflow boundary conditions as
    closely as possible with patient-specific ones. As proposed by
    \citet{Valen-Sendstad2015}, it was computed as follows:
    \begin{equation} \label{eq:scaledFlowRate}
        \gls{patientBloodFlowRate}
        \functionOf{\gls{time}}
        =
        \gls{timeBloodFlowRate}
        \functionOf{\gls{time}}
        \left(
            \frac{
                \gls{arteryDiameter}
            }{
                \gls{populationAvgDiameter}
            }
        \right)^2\,,
    \end{equation}
    where \gls{arteryDiameter} is the diameter of the patient's \gls{ica} at
    the inlet (see \cref{fig:boundaryConditions}a) and the square power was
    recommended by \citet{Valen-Sendstad2015} for cerebral arteries --- rather
    than the cube power predicted by Murray's law. The population-average
    diameter of the \gls{ica}, \gls{populationAvgDiameter}, was the sample's
    \gls{ica} average diameter at the truncated location, yielding a value of
    \SI{4.35}{\milli\meter}. The populational-averaged flow rate waveform,
    \gls{timeBloodFlowRate}, was selected according to the patient's age. The
    profiles measured at the cerebral arteries for \enquote{older adults} by
    \citet{Hoi2010}, who measured blood flow rates in patients with an average
    age of 68 $\pm$ 8 years, and for \enquote{young adults} by
    \citet{Ford2005}, who conducted similar measurements in patients with an
    average age of 28 $\pm$ 7 years (see \cref{fig:boundaryConditions}b).
    Based on these thresholds, the sample cases were categorized into
    \enquote{older adults}, age $\ge$ \num{50} years, and \enquote{young
    adults}, age $<$ \num{50} years. The threshold of \num{50} years was
    selected as an approximation of the average age separating these two
    groups.

    Additionally, zero-pressure gradient was applied at the inlet. At the
    vascular outlets, a flux-corrected zero-gradient condition was employed for
    the velocity, and a resistance condition, where the outlet pressure is
    proportional to the flow profile over time \citep{Chnafa2018}, was applied
    for the pressure (\cref{fig:boundaryConditions}b).

\subsection{%
    Numerical Strategies%
}   \label{sec:numericalStrategiesRt1}

    \Cref{eq:continuityEquation,eq:momentumEquation} were numerically solved
    using OpenFOAM\R (v23.12). To maintain second-order accuracy, we employed a
    second-order upwind scheme for advective terms, the Green-Gauss scheme for
    velocity and pressure gradients, and second-order central differences for
    diffusion terms --- incorporating non-orthogonal and skewness corrections
    \citep{JasakThesis1996}. Pressure-velocity coupling was managed via the
    \gls{piso} algorithm \citep{Issa1986}.

    Computational meshes were generated with \command{snappyHexMesh}, resulting
    in predominantly hexahedral cells. The mesh conforms to the curved lumen
    boundary using polyhedral cells at the wall, supported by a five-layer
    prismatic boundary-layer refinement. Elements were designed to refine
    progressively toward the wall. Based on extensive mesh and time-step
    sensitivity studies across various \glspl{ia} geometries, we adopted an
    optimal cell density of, at least, \SI{3000}{cells\per\milli\cubic\meter}
    and the temporal discretization utilized a fixed time-step of
    \SI{1e-04}{\second} \citep{Oliveira2021a}. To eliminate initial transient
    effects, three cardiac cycles were simulated, with only the final cycle
    retained for hemodynamic analysis.

\subsection{%
    Data Analysis%
}   \label{sec:dataAnalysis}

\subsection*{Subdivisions of the Aneurysm Sac Surface by Local Shape}

\begin{figure}[!t]
        \caption{%
            Example, for case C0001 of the Aneurisk repository, of the (a)
            Gaussian, \gls{GaussianCurvature}, curvature fields of the lumen
            surface, and (b) the resulting local-shape patching, based on the
            one proposed by \citet{Ma2004}, depicted at the bottom of the panel
            in a tabular format, with an illustration of what shape each
            surface has, and (c) the neck, body, and dome patches (figure
            adapted from \citet{Oliveira2023}).
        }

        \includegraphics[width=\textwidth]{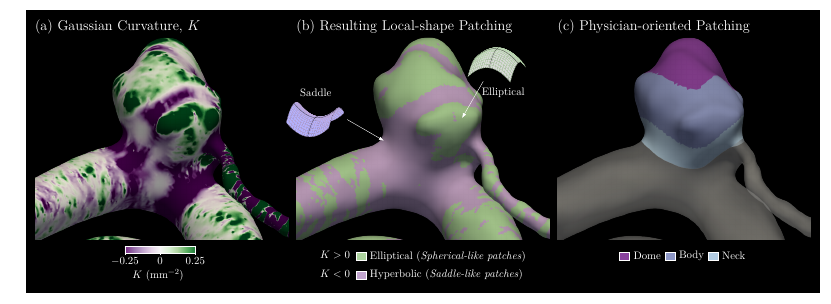}

        \label{fig:sacShapePatching}
    \end{figure}

    The aneurysm sac surface was partitioned using two patching schemes. First,
    we used the Gaussian curvature (\gls{GaussianCurvature}) to classify
    surface topology \citep{Ma2004,Oliveira2023}. At any given point on a
    surface, the Gaussian curvature is defined as the product of the two
    principal curvatures that represent the maximum and minimum normal
    curvatures at that point. Gaussian curvature fields, measured in
    \si[per-mode=reciprocal]{\per\square\milli\meter}, were computed in
    \gls{vtk}\R and classified with an in-house Python code (see
    \cref{fig:sacShapePatching}a for an example on case C0001 of the Aneurisk
    database). Gaussian curvature categorizes regions as elliptic
    ($\gls{GaussianCurvature} > 0$, ellipsoidal), parabolic
    ($\gls{GaussianCurvature} = 0$, cylindrical), or hyperbolic
    ($\gls{GaussianCurvature} < 0$, saddle-shaped) (see
    \cref{fig:sacShapePatching}b where we included a detail of a typical shape
    of saddle and elliptical surfaces). Our mapping focused on morphologically
    relevant classes found in the vascular sample \citep{Ma2004} --- saddle and
    elipsoidal-shaped ---, excluding physiologically improbable types like
    perfectly planar surfaces that were not identified. For simplicity, in our
    discussions, we refer to elliptic regions as \enquote{spherical-like} and
    hyperbolic regions as \enquote{saddle-like}.

    Second, the sac was divided into clinically relevant regions:
    \enquote{neck}, \enquote{body}, and \enquote{dome}. Following
    \citet{SalimiAshkezari2021}, we calculated the geodesic distance from the
    neck contour, \gls{distanceToNeckLine}, using \gls{vtk}\R. The aneurysm sac
    surface was partitioned based on the maximum geodesic distance,
    \gls{distanceToNeckLineMax}, and categorized as \emph{neck} for
    \gls{distanceToNeckLine} up to $0.20\gls{distanceToNeckLineMax}$,
    \emph{body} for values between $0.20\gls{distanceToNeckLineMax}$ and
    $0.60\gls{distanceToNeckLineMax}$, and the \emph{dome} for distances
    exceeding $0.60\gls{distanceToNeckLineMax}$ (see
    \cref{fig:sacShapePatching}c).

\subsection{Hemodynamics and Flow Characterization}

    From the \gls{cfd} results, our analysis proceeded in two parts. First, we
    focused on hemodynamic quantities derived from the \gls{wss} vector,
    \gls{wallShearStressVector} (bold-face font for vectors). We computed the
    \gls{tawss}, defined as the temporal average of the \gls{wss} magnitude,
    \gls{wallShearStressMag} (no bold-face font for magnitudes):
    \begin{equation} \label{eq:timeAvgWallShearStress}
        \gls{timeAvgWallShearStress}
        =
        \temporalavg{\gls{cardiacPeriod}}{
            \gls{wallShearStressMag}
        }\,,
    \end{equation}
    where \gls{cardiacPeriod} is the cardiac period. Additionally, we
    calculated the \gls{osi}, which quantifies the temporal variation in the
    direction of the \gls{wss} vector \citep{He1996}:
    \begin{equation} \label{eq:oscillatoryShearIndex}
        \gls{oscillatoryShearIndex}
        =
        \frac{1}{2}
        \left(
            1
            -
            \frac{
                \left\|
                    \temporalavg{\gls{cardiacPeriod}}{%
                        \gls{wallShearStressVector}
                        \functionOf{
                            \gls{EulerCoord},
                            \gls{time}
                        }
                    }
                \right\|
            }{
                \temporalavg{\gls{cardiacPeriod}}{%
                    \norm{
                        \gls{wallShearStressVector}
                        \functionOf{
                            \gls{EulerCoord},
                            \gls{time}
                        }
                    }
                }
            }
        \right)\,.
    \end{equation}
    We also computed and analyzed the \gls{tawssg} as defined in
    \citet{Geers2017} to quantify spatial gradients of \gls{tawss}. Its
    definition and calculation are detailed in the Supplementary Material.

\begin{figure}[!b]
        \caption{
            Schematic representation of the fixed points topology of the
            cycle-averaged \gls{wss} vector field. Arrows indicate lines of
            flow separation or impingement originating from or terminating at
            these points (inspired by figures in
            \citet{Surana2006,Mazzi2021}).
        }

        \includegraphics[width=\textwidth]{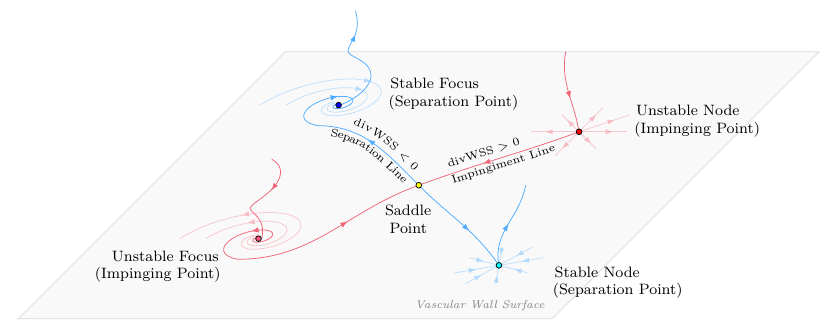}

        \label{fig:fixedPointsTopology}
    \end{figure}

    Second, to characterize intra-aneurysmal flow and its impact on the wall,
    we identified fixed points of the cycle-averaged \gls{wss} vector field
    \citep{Surana2008}. In this context, fixed points --- namely, saddles, foci
    and nodes --- are locations where the \gls{wss} vector vanishes,
    effectively representing stagnation points where the near-wall velocity is
    zero \citep{Gambaruto2012,Arzani2018a}. These points mark the origin or
    termination of separation and impingement lines \citep{Surana2006},
    referred to as \emph{flow separatrices}. These trajectories were computed
    by integrating the cycle-averaged \gls{wss} vector field on the lumen
    surface \citep{Mazzi2020}. \cref{fig:fixedPointsTopology} illustrates some
    fixed point topologies observed in this study, although other topologies
    may occur. For the sake of clarity, we adopt the terms \enquote{separation}
    and \enquote{impingement} rather than standard dynamical systems
    nomenclature, as can be seen in \cref{fig:fixedPointsTopology}.

    Complementing this, we used an Eulerian approach based on the divergence of
    the normalized cycle-averaged \gls{wss} field to identify regions of flow
    approaching or leaving the wall \citep{Wu2006,Surana2006,Mazzi2020}:
    \begin{equation}
        \gls{wssDivergence}
        \equiv
        \div{
        \left(
            \dfrac{
                \tavg{\gls{wallShearStressVector}}
            }{
                \norm{
                    \tavg{\gls{wallShearStressVector}}
                }
            }
        \right)
        },.
    \end{equation}

    Physically, positive \gls{wssDivergence} characterizes impingement zones
    where flow directed toward the wall is forced to spread laterally occurring
    around a impingiment line. Conversely, negative \gls{wssDivergence}
    identifies separation regions, capturing regions where the flow is leaving
    the luminal surface and around a separation line, as illustrated in
    \cref{fig:fixedPointsTopology}.

\subsection{Statistical Analysis}

    Statistical analysis was conducted on $n=76$ aneurysm cases using the SciPy
    library \citep{scipy} with a significance level of $\gls{significanceLevel}
    = 0.05$. Normality was assessed via the Shapiro-Wilk test. To compare
    metrics between ruptured and unruptured cases, lateral and bifurcation
    aneurysms, and across different patch types, we employed paired t-tests for
    normally distributed data and Wilcoxon signed-rank tests for non-normal
    distributions.

    The statistical analysis was performed on the surface-averaged metric for
    the \gls{tawss}, \gls{osi}, and \gls{wssDivergence} across each patch. The
    surface average of a field $F$ was defined as:
    \begin{equation} \label{eq:surfaceAvg}
        \savg{F}_{\gls{surface}_i}
        =
        \surfaceavg{
            \gls{surface}_i
        }{
            F
        }\,,
    \end{equation}
    where $A(\gls{surface}_i)$ denotes the surface area of the specific
    saddle-like or spherical-like patches identified in
    \cref{fig:sacShapePatching}b.

    \section{Results} \label{sec:results}

\subsection*{Aneurysm Sac Patches Characterization}

\begin{figure}[!htb]
        \caption{%
            Distributions of (a) surface area percentage of curvature patches
            relative to the total aneurysm area and (b) surface-averaged
            Gaussian curvature across neck, body, and dome regions.
        }

        \includegraphics[width=\textwidth]{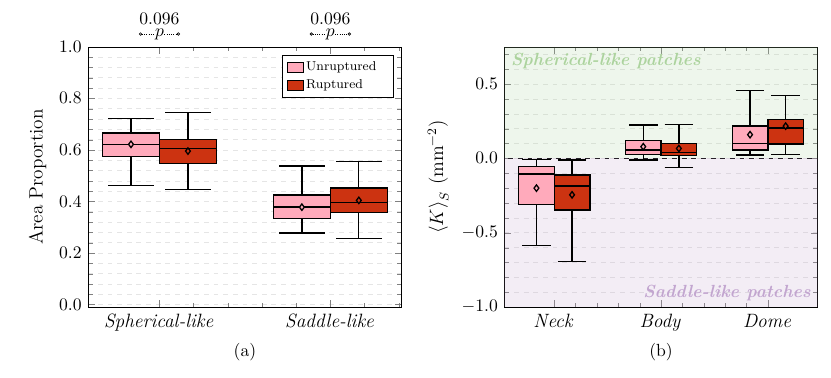}

        \source{prepared by the author}
        \label{fig:boxPlotsPhysicianOrientedPatching}
    \end{figure}

    To evaluate the morphological composition of the aneurysm wall, we
    categorized the surface into discrete patches based on local curvature
    properties. This section details the prevalence and spatial distribution of
    these patches across the aneurysm sac, as summarized in
    \cref{fig:boxPlotsPhysicianOrientedPatching}. Spherical-like patches are
    more prevalent on \gls{ia} sacs, while saddle-like patches occupy
    \SIrange{30}{45}{\percent} of the surface area and spherical-like are, on
    average, \SI{56.7}{\percent} larger ($\gls{pValue} < 0.001$). These area
    distributions do not differ significantly between ruptured and unruptured
    aneurysms (see $p$-values in
    \cref{fig:boxPlotsPhysicianOrientedPatching}a).

    The distributions of \savg{\gls{GaussianCurvature}} for physician-oriented
    patching (\cref{fig:boxPlotsPhysicianOrientedPatching}b) reveal distinct
    morphological patterns. Neck patches are characterized by negative Gaussian
    curvature, identifying them as predominantly saddle-like. In contrast,
    aneurysm domes are primarily composed of spherical-like patches, whereas
    the body patches exhibit intermediate morphological characteristics between
    these two extremes. This pattern can be noticed by the example of case
    C0001 comparing \cref{fig:sacShapePatching}a and c.

\subsection*{Local Hemodynamics on Saddle-like and Spherical-like Patches}

\begin{figure}[!b]
        \caption{%
            Distributions of (a) \savg{\gls{timeAvgWallShearStress}} and (b)
            \savg{\gls{oscillatoryShearIndex}} across spherical-like and
            saddle-like patches, with the sample segregated by rupture status.
        }

        \includegraphics[width=\textwidth]{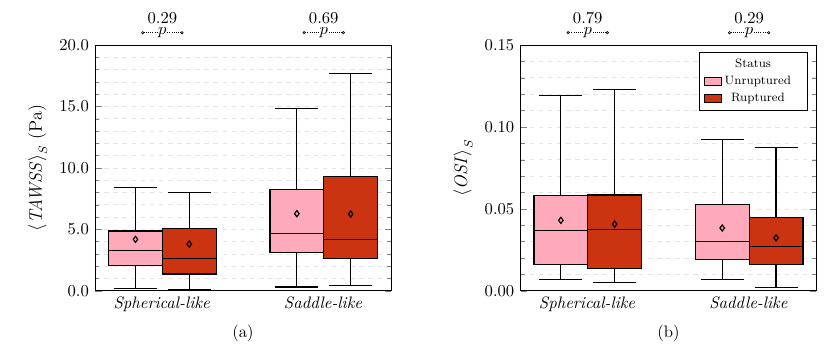}

        \source{prepared by the author}
        \label{fig:boxPlotsLocalHemodynamics}
    \end{figure}

    Hemodynamic data from the Aneurisk repository reveal that \gls{tawss} was
    significantly higher on saddle-like patches compared to spherical-like
    regions ($\gls{pValue} \ll \num{0.001}$,
    \cref{fig:boxPlotsLocalHemodynamics}a), with an average increase of
    \SI{55.7}{\percent}. Conversely, \gls{osi} exhibited the opposite trend,
    averaging \SI{15.0}{\percent} lower on saddle-like patches ($\gls{pValue}
    \ll \num{0.001}$, \cref{fig:boxPlotsLocalHemodynamics}b). Notably, the
    difference in \gls{rrs} --- a parameter correlated with \gls{osi} that
    identifies local flow recirculation near the lumen --- was even more
    pronounced; saddle-like patches presented an average \gls{rrs}
    \SI{95.5}{\percent} lower than spherical-like patches. Furthermore, the
    \gls{timeAvgWallShearStressGradient} showed significantly higher dispersion
    on saddle-like regions, with a standard deviation \num{3.4} times larger
    than on spherical-like patches (data provided in Supplementary Material).
    This suggests that \gls{tawss} on spherical-like surfaces is more spatially
    uniform along the flow direction than on saddle-like surfaces. No
    significant differences were observed in any hemodynamic parameters when
    comparing saddle-like or spherical-like patches between ruptured and
    unruptured subgroups ($p$-values shown in
    \cref{fig:boxPlotsLocalHemodynamics}). Similarly, no significant variations
    were detected when comparing lateral and bifurcation cases within each
    patch type (data not shown).

\begin{figure}[!b]
        \caption{%
            Local hemodynamics in a representative lateral aneurysm (case
            C0002): (a) local shape patching, (b) \gls{tawss} field, and (c)
            \gls{osi} field (the yellow line marks the aneurysm neck).
        }

        \includegraphics[width=\textwidth]{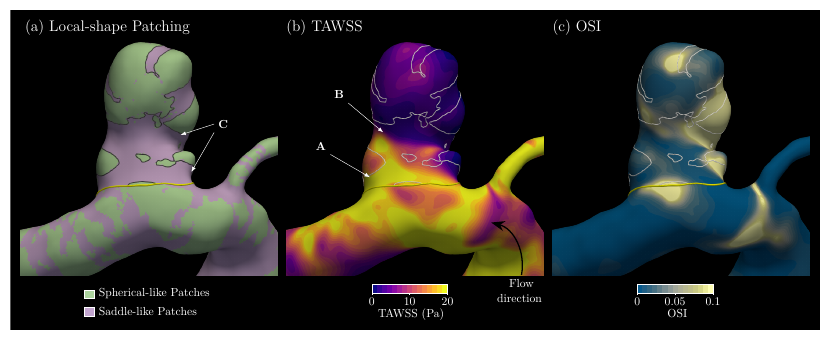}

        \source{prepared by the author}
        \label{fig:hemodynamicLateralCase}
    \end{figure}

    A representative example of these hemodynamic patterns is seen in case
    C0002 (\cref{fig:hemodynamicLateralCase}), a lateral aneurysm with a
    balanced distribution of saddle-like and spherical-like regions. As shown
    in \cref{fig:hemodynamicLateralCase}b, \gls{tawss} is high near the neck
    (point A), characterized by saddle-like geometry, but decreases as the flow
    enters the sac and passes through another saddle-like region (point B).
    This results in reduced \gls{tawss} within the spherical-like region of the
    dome. In contrast, the \gls{osi} field (\cref{fig:hemodynamicLateralCase}c)
    is elevated in the dome, coinciding with spherical-like patches, while
    lower values are maintained on saddle-like patches near the neck and body.

\subsection*{Correlation between Hemodynamics and Wall Curvatures}

\begin{figure}[!t]
        \caption{%
            Correlation between \savg{\gls{timeAvgWallShearStress}} (a, b) and
            \savg{\gls{oscillatoryShearIndex}} (c, d) versus
            \savg{\gls{GaussianCurvature}} across spherical and saddle-like
            patches, segregated by rupture status. (Spearman's correlation
            coefficients and p-values annotated were computed with SciPy)
        }

        \includegraphics[width=\textwidth]{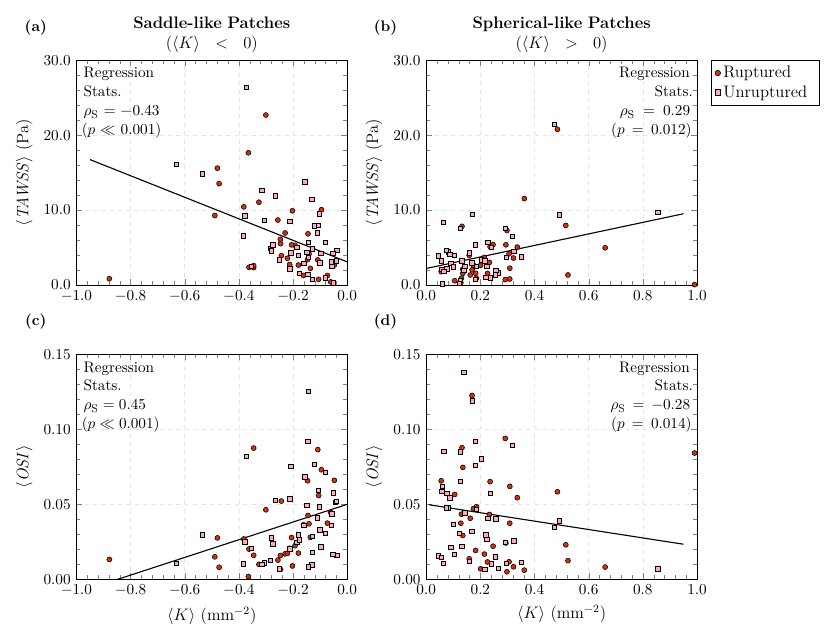}

        \source{prepared by the author}
        \label{fig:corrHemoVarVsCurvatures}
    \end{figure}

    The definitions of spherical-like and saddle-like patches are rooted in the
    Gaussian curvature of the lumen. The results in
    \cref{fig:boxPlotsLocalHemodynamics} suggest a correlation between local
    hemodynamics and these morphological metrics. Our analysis confirms a
    distinct correlation between selected hemodynamic parameters and Gaussian
    curvature, as the data in \cref{fig:corrHemoVarVsCurvatures} reveals.
    Specifically, regions with increasingly negative Gaussian curvature tend to
    exhibit higher \gls{tawss}, lower \gls{osi}
    (\cref{fig:corrHemoVarVsCurvatures}a and c), and higher, more positive
    \gls{tawssg} (see Supplementary Material). These conditions suggest that on
    saddle-like patches, high and growing \gls{tawss} occurs in conjunction
    with a constant \gls{wss} vector direction that does not undergo
    significant rotation during the cardiac cycle. Areas of pronounced negative
    Gaussian curvature correspond to surfaces that are more sharply curved in
    the two principal directions, as exemplified by the \enquote{squeezed}
    saddle topology at region C in \cref{fig:hemodynamicLateralCase}.

    These correlations invert on spherical-like patches. In these regions,
    \gls{tawss} increases and \gls{osi} decreases as curvature increases,
    though the sensitivity of \gls{osi} to curvature is less pronounced than on
    saddle-like patches (\cref{fig:corrHemoVarVsCurvatures}b and d). While
    \gls{tawssg} is slightly positively correlated with curvature on
    spherical-like surfaces, its average remains much closer to zero and is
    statistically distinct from the values observed on saddle-like patches
    (see Supplementary Material).

\subsection*{Wall Adjacent Flow Causing Hemodynamic Patterns}

    To interpret these hemodynamic patterns, it is essential to identify the
    underlying 3D flow structures adjacent to the lumen. High \gls{wss} regions
    are fundamentally linked to high wall vorticity, \gls{vorticity}, as
    described by the relationship at at no-slip boundaries \citep{Wu2006}:
    \begin{equation}
        \gls{wallShearStressVector}
        =
        \gls{dynamicViscosity}
        \left(
            \gls{vorticity}
            \cprod
            \gls{sNormalVector}
        \right)
    \end{equation}
    where \gls{sNormalVector} is the outward unit normal vector. Additionally,
    given that vortices are typically generated in regions of flow impingement
    and separation, the \gls{wss} vector field on an aneurysm sac is rarely
    unidirectional. Consequently, \gls{osi} levels
    (\cref{eq:oscillatoryShearIndex}) tend to be higher where the adjacent flow
    departs from a unidirectional path. To determine if this behavior
    predominates over spherical-like patches, we examined markers of flow
    separation and attachment.

\begin{figure}[!b]
        \caption{%
            Flow characterization in case C0002: (a) curvature patch type; (b)
            \gls{wssDivergence} with surface limiting streamlines (of the
            cycle-averaged \gls{wss} vector field with the line-integral
            convolution technique), fixed points, and separatrix lines; (c)
            cycle-averaged \gls{lambda2VortexId} field on the vascular surface
            (the yellow line marks the aneurysm neck).
        }

        \includegraphics[width=\textwidth]{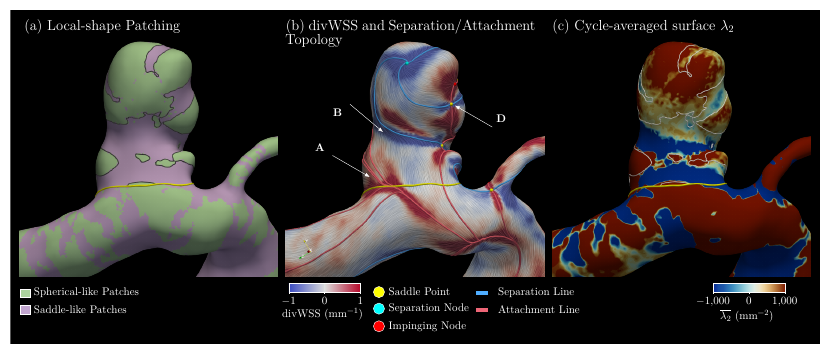}

        \label{fig:limitingSurfaceFlowCharacterization}
    \end{figure}

    Using case C0002 as a representative example
    (\cref{fig:limitingSurfaceFlowCharacterization}), we observe flow
    originating from the \gls{ica} sweeping across the aneurysm neck. This
    generates a region of positive \gls{wssDivergence} and high \gls{tawss}
    (point A in \cref{fig:hemodynamicLateralCase}b), where an attachment line
    crosses the neck (\cref{fig:limitingSurfaceFlowCharacterization}b). As the
    flow enters the sac, it separates from the wall at a separation line
    terminating at a separation point (stable node) near the dome. These
    regions are marked by negative \tavg{\gls{lambda2VortexId}} values,
    indicating intense near-wall vorticity responsible for the elevated
    \gls{tawss} (\cref{fig:limitingSurfaceFlowCharacterization}c; this can also
    be confirmed in the Supplementary Material where we present the temporal
    evolution of the \gls{lambda2VortexId} field and the \gls{wss} with a video
    and a figure).

\begin{figure}[!htb]
        \centering
        \caption{%
            Distribution, measured by the probability of occurrence, of fixed
            point types across different curvature patches.
        }
        \includegraphics[width=\textwidth]{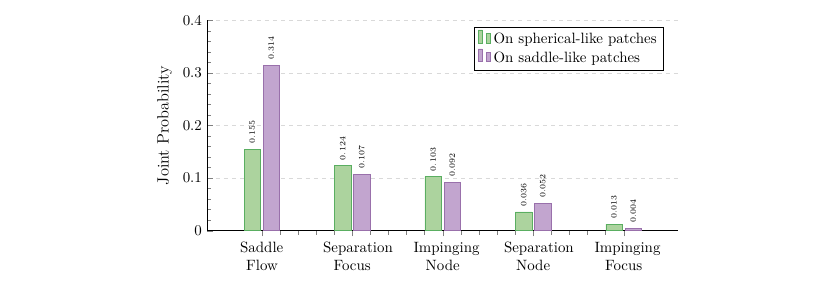}

        \label{fig:fixedPointsDistribution}
    \end{figure}

    This pattern is consistent across the sample. In lateral aneurysms, the
    inflow typically surrounds the wall, forming a vortex at a separation focus
    or node, often on the dome. This stream may impinge on the dome (see
    impinging node on the region of positive \gls{wss} divergence in
    \cref{fig:limitingSurfaceFlowCharacterization}b) before detaching at a
    separation line that originates at a saddle point near the neck or body (an
    example can be seen in point D in
    \cref{fig:limitingSurfaceFlowCharacterization}b). In bifurcation aneurysms,
    the behavior is more complex and, by inspection of the bifurcation cases in
    the sample, seems to depend on the aneurysm orientation relative to the
    parent artery. However, inflow jets frequently impinge on the dome,
    producing an impinging node and a subsequent separation line where the jet
    meets the circulating flow. The relationship between these flow features
    and the underlying wall morphology are supported by the distribution of
    fixed point types, where saddle points, separation foci, and impinging
    nodes are the most prevalent features as can be seen in
    \cref{fig:fixedPointsDistribution}, showing the joint probability
    distribution of the fixed points. The joint probability represents the
    likelihood of a specific fixed-point type occurring concurrently with a
    specific curvature patch, effectively identifying a topological flow
    structure to the local geometry.

\begin{figure}[!t]
        \caption{%
            (a) Distribution of impingement and separation line lengths by
            patch type. Only aneurysmal flow features are included. (b) Box
            plots of the \gls{wssDivergence} distributions by patch type and
            rupture status.
        }

        \includegraphics[width=\textwidth]{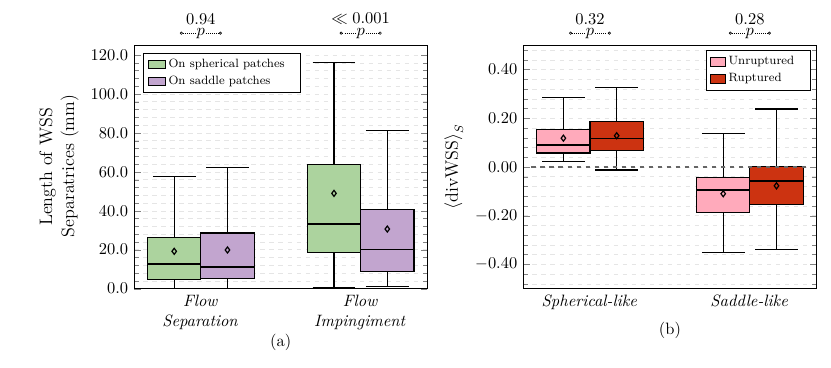}

        \label{fig:boxPlotsLengthWSSSeparatrices}
    \end{figure}

    Analysis of separatrix lengths (see
    \cref{fig:boxPlotsLengthWSSSeparatrices}a) further reveals that
    spherical-like patches are more likely to exhibit longer impingement lines
    than saddle-like patches. Specifically, on spherical-like patches,
    impingement lines are approximately \SI{155}{\percent} longer than
    separation lines. This aligns with the observed positive \gls{wss}
    divergence on these patches (see
    \cref{fig:boxPlotsLengthWSSSeparatrices}b), indicating that impinging flow
    dominates the more spherical dome regions. The prevalence of these
    structures explains the higher \gls{osi} values observed in spherical-like
    regions, where the instantaneous \gls{wss} vector fluctuates significantly
    due to the impinging flow.

    Impingement lines are also significantly longer on saddle-like patches
    ($\gls{pValue} \ll \num{0.001}$) by \SI{54.0}{\percent}, on average.
    Indeed, regions of positive \gls{wssDivergence} are prevalent on these
    patches (\cref{fig:boxPlotsLengthWSSSeparatrices}b), supporting our
    previous observation that the sweeping inflow stream first encounters
    saddle-like geometry near the neck. Nevertheless, negative \gls{wss}
    divergence --- a marker of flow moving away from the wall --- is also
    common on saddle-like patches. This is likely explained by regions of
    vortex formation that coincide with negative \gls{lambda2VortexId} values,
    as demonstrated in case C0002
    (see \cref{fig:limitingSurfaceFlowCharacterization}c). Furthermore, the
    regions of higher \gls{tawss} coincide with negative \gls{lambda2VortexId}
    (comparing \cref{fig:hemodynamicLateralCase}b and
    \cref{fig:limitingSurfaceFlowCharacterization}c). Previous works have
    established a relationship between near-boundary vortical activity and
    \gls{wss} peaks. \citet{ElHassan2012} found that vortical structures
    advected from a jet shear layer correlated with the location of \gls{wss}
    peaks at the wall. Similarly, \citet{Wild2023} identified hairpin vortices
    in the carotid bifurcation responsible for high wall shear stress.
    Therefore, it is reasonable to conclude that the elevated \gls{tawss} on
    saddle-like patches is primarily due to the formation of vortices adjacent
    to these regions, identified by negative \gls{lambda2VortexId} values in
    areas of detaching flow.

\begin{figure}[!t]
        \centering
        \caption{%
            Correlation plot between surface-averages of \gls{lambda2VortexId}
            and \gls{wssDivergence} categorized by patch type (Spearman's
            correlation coefficient and p-value annotated were computed with
            SciPy).
        }

        \includegraphics[width=\textwidth]{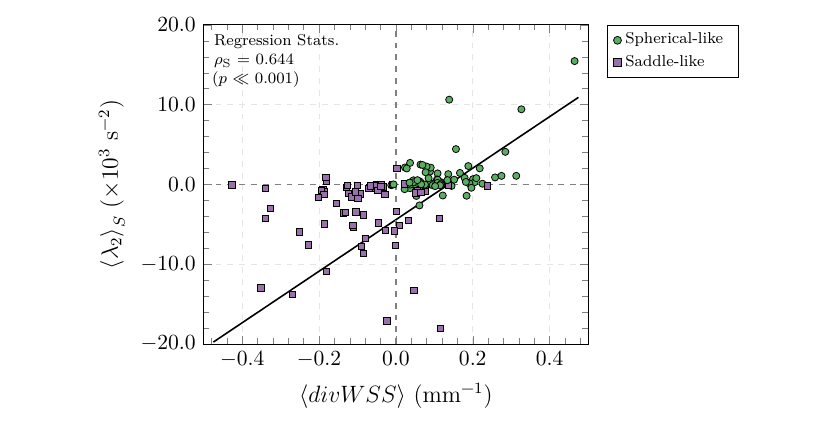}

        \label{fig:plotCorrDivWSSVsLambda2}
    \end{figure}

    Finally, regions of negative \gls{lambda2VortexId} are most commonly found
    on hyperbolic (saddle-like) patches as supported by our data
    (\cref{fig:plotCorrDivWSSVsLambda2}), where the correlation between
    surface-averaged \gls{lambda2VortexId} and \gls{wssDivergence} is
    statistically significant ($\gls{pValue} \ll \num{0.001}$) for both patch
    types. Most importantly, the results indicate that spherical-like patches
    tend to be adjacent to flow with positive \gls{lambda2VortexId} and
    positive \gls{wssDivergence}; here, the flow is dominantly impinging with
    low \gls{tawss} due to minimal near-wall vortical activity. In contrast,
    flow on saddle-like patches is more likely to exhibit negative average
    \gls{lambda2VortexId} and high-intensity vorticity, marking regions of
    detaching flow and elevated shear stress.

    \section{Discussion}
\label{sec:discussion}

    Our findings elucidate critical aspects of hemodynamics within a relatively
    large cohort of \glspl{ia}. First, our results corroborate the hypothesis
    proposed by \citet{Oliveira2023}: a clear relationship exists between
    luminal curvature and local hemodynamics that is maintained across a broad
    sample. Notably, these correlations persist across sub-samples categorized
    by rupture status and aneurysm type (lateral or bifurcation). This is
    particularly significant because inflow patterns differ substantially
    between these types; lateral aneurysms typically feature a single main
    vortex, while bifurcation aneurysms exhibit complex patterns with multiple
    vortices. Thus, the relationship between curvature and hemodynamics appears
    robust to variations in global flow topology. Furthermore, these findings
    were obtained using patient-specific boundary conditions, with flow rates
    scaled by vessel diameter and patient age, further strengthening the
    results.

    Our results align with previous studies characterizing the hemodynamic
    environments of intracranial aneurysms. In a large-scale analysis,
    \citet{Karnam2024} reported low \gls{wss} and high \gls{osi} on the
    aneurysm dome, contrasted by high \gls{wss} and low \gls{osi} primarily at
    the neck. While their study did not explicitly incorporate wall curvature,
    their anatomical findings are consistent with our geometric framework,
    given that the dome is predominantly composed of spherical-like patches,
    whereas the neck consists mainly of saddle-like patches. Furthermore,
    \citet{Varble2017} investigated vortical activity within the aneurysm sac
    using the Q-criterion \citep{Wu2006} --- a method analogous to the
    \gls{lambda2VortexId}-criterion employed in our analysis. They observed
    that high \gls{wss} values occurred in regions of positive Q, which
    identify vortical structures adjacent to the wall, further supporting our
    observation that near-wall vorticity drives elevated shear stress.

    Abnormal hemodynamic conditions are known to trigger biological cascades in
    vessel tissue. Specifically, the inception of \glspl{ia} has been linked to
    elevated \gls{wss} and high, positive \gls{tawssg}
    \citep{Dolan2013,Frosen2019}. These conditions occur naturally at vascular
    bifurcations due to flow impingement, where the apex is characterized by
    saddle-like points. \citet{Frosen2019} described an \enquote{inflammation-
    mediated} pathway for \gls{ia} inception, highlighting a feedback loop that
    may continue even if \gls{wss} reestablishes baseline levels. Our findings
    suggest that \gls{wss} may remain elevated due to the saddle-like nature of
    the initial bulge at the neck, potentially reinforcing this inflammatory
    pathway.

    While there is consensus on \gls{ia} inception, the mechanisms governing
    subsequent growth remain a subject of debate. \citet{Meng2014} proposed a
    \enquote{high-flow} pathway, where high \gls{wss}, positive gradients, and
    low \gls{osi} degrade the wall matrix, leading to thin-walled blebs.
    Conversely, a \enquote{low-flow} environment (low \gls{tawss} and high
    \gls{osi}) may trigger an inflammatory response leading to thick,
    atherosclerotic walls --- the \enquote{low-flow} pathway
    \citep{Furukawa2018,Cebral2019}. Our results suggest preferential locations
    for these events: saddle-like (hyperbolic) patches favor high-flow
    conditions, while spherical-like (elliptic) patches favor low-flow
    conditions.

    Furthermore, using the terminology of \citet{Meng2014}, high-flow
    conditions are often assumed to be synonymous with impinging flow, while
    low-flow conditions are linked to slow recirculation. Our results suggest
    this assumption is oversimplified. High-flow regions can occur
    independently of impinging jets, such as near flow detachment points after
    passing through a hyperbolic patch, as seen in
    \cref{fig:hemodynamicLateralCase}. While both impingement and detachment
    can occur on saddle-like patches, spherical-like patches are predominantly
    sites of impingement. This indicates that wall curvature, independent of
    global flow patterns, is the primary determinant of local hemodynamic
    behavior.

\begin{figure}[!t]
        \caption{%
            (a) Area proportion of abnormal-hemodynamic patches relative to the
            total aneurysm sac surface area; (b) Distribution of
            surface-averaged Gaussian curvature per abnormal-hemodynamics
            patches of type I (red-thin), II (white-thick), and also
            \enquote{regular} --- neither Type-I nor Type-II.
        }

        \includegraphics[width=\textwidth]{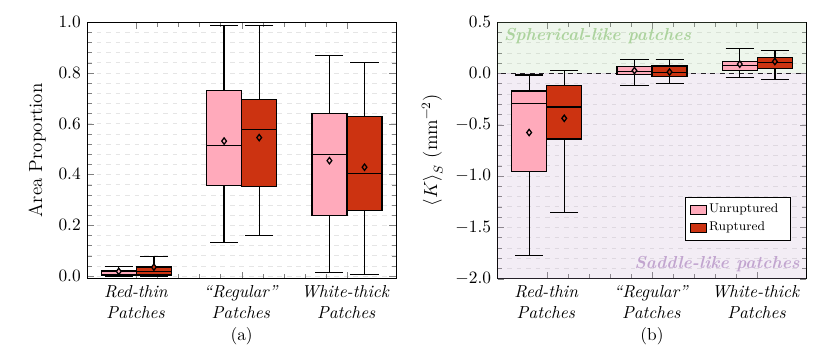}

        \source{prepared by the author}
        \label{fig:boxPlotGaussCurvPerWallType}
    \end{figure}

    \citet{Meng2014} hypothesized that these pathways lead to distinct wall
    phenotypes: \enquote{Type-II} (thick, atherosclerotic, yellow-whitish) and
    \enquote{Type-I} (thin, reddish). We investigated these phenotypes using
    the hemodynamic thresholds from our previous work \citep{Oliveira2023},
    where low-flow patches were identified by $\gls{tawss} < \SI{5}{\pascal}$
    and $\gls{osi} > \SI{0.01}{\pascal}$, and high-flow regions by $\gls{tawss}
    > \SI{10}{\pascal}$ and $\gls{osi} < \SI{0.001}{\pascal}$. These thresholds
    were established based on the averaged values reported by
    \citet{Furukawa2018} and \citet{Cebral2019} (see
    \cref{fig:boxPlotGaussCurvPerWallType}). Notably, white-thick (Type-II)
    patches were significantly more prevalent than red-thin (Type-I) patches
    across our sample ($\gls{pValue} < \num{0.001}$), irrespective of rupture
    status.

    Excluding extreme outliers, saddle-like patches are more susceptible to the
    hemodynamic conditions favoring red-thin (Type-I) phenotypes, whereas
    spherical-like patches correlate with thick, atherosclerotic (Type-II)
    phenotypes. This aligns with findings suggesting that thin walls correlate
    with low \gls{osi} \citep{Kimura2019}. If validated in longitudinal
    studies, these relationships could facilitate the early identification of
    regions prone to specific wall phenotypes. Such information would be
    important for endovascular planning, as Type-I patches are thinner and
    significantly more susceptible to intraprocedural rupture
    \citep{Chen2016,Chen2017}.

    \subsection*{Limitations}

    Despite the robust correlations observed in this cohort, certain
    limitations merit discussion. First, Gaussian curvature calculations are
    inherently sensitive to numerical noise in discretized surfaces
    \cite{Ma2004}. To minimize these errors, we utilized highly-refined surface
    meshes for all geometric computations. Furthermore, as our conclusions are
    derived from broad morphological regions and surface-averaged metrics
    rather than point-wise values, it is unlikely that localized discretization
    noise influenced the primary findings of this study.

    Second, regarding the physical modeling of blood flow, we assumed rigid
    vessel walls. While wall compliance and the resulting fluid-structure
    interaction can influence near-wall hemodynamics, the rigid-wall assumption
    is a well-established and validated standard for numerical investigations
    of intracranial aneurysms \cite{Cebral2005}, particularly when focusing on
    time-averaged flow patterns. Future research incorporating moving
    boundaries may further elucidate the secondary effects of wall motion on
    the curvature-hemodynamic relationship established here.

    Finally, the thresholds used to define low-flow and high-flow hemodynamics
    analyzed in \cref{fig:boxPlotGaussCurvPerWallType}b are relatively
    arbitrary. As detailed in our prior work \citep{Oliveira2023}, these values
    were synthesized from existing literature. To mitigate selection bias, we
    performed a parametric sensitivity study by independently varying the
    \gls{tawss} and \gls{osi} thresholds. The fundamental structure of the
    distributions remained unchanged: saddle-like patches consistently favored
    high-flow conditions, while spherical-like patches favored low-flow
    environments, regardless of the specific cutoff values. This stability
    confirms that the relationship between curvature and hemodynamics is robust
    to threshold selection, further strengthening the conclusions of this
    study.

\section{Conclusions}

    This study establishes local luminal curvature as a primary and independent
    determinant of the hemodynamic environment in \glspl{ia}. By analyzing a
    large, heterogeneous cohort, we demonstrated that the relationship between
    surface morphology and hemodynamics remains robust, regardless of aneurysm
    type, location, or rupture status. Specifically, we show that saddle-like
    patches --- typically found at the aneurysm neck --- are preferentially
    associated with elevated \gls{tawss}, low \gls{osi}, driven by intense
    near-wall vortical activity. Conversely, spherical-like regions,
    predominant at the dome, act as sites of flow impingement characterized by
    lower shear stress and higher \gls{osi}.

    A key contribution of this work is the mechanistic link established between
    these geometric features and specific pathological wall phenotypes:
    saddle-like regions are prone to thin, reddish phenotypes, while
    spherical-like regions favor thick, atherosclerotic developments. These
    findings underscore the clinical potential of curvature-based mapping as a
    powerful, non-invasive tool. By leveraging readily available morphological
    data from routine imaging, this approach enables superior risk
    stratification --- particularly when integrated into objective clinical
    scoring systems --- more accurate prediction of wall vulnerability, and
    enhanced precision in planning endovascular interventions.

    \section{Acknowledgements}

    This work was supported by resources supplied by the Center for Scientific
    Computing (NCC/GridUNESP) of the \gls{unesp}
    (\url{www2.unesp.br/portal#!/gridunesp}) and by the National Laboratory of
    Scientific Computing through the use of the SDumont supercomputer (projet
    ID CFDIA). Funding: This research was supported by grants 2023/06609-9 and
    2025/23522-0 of \gls{fapesp}.

    \bibliographystyle{elsarticle-num-names}
    \bibliography{references}

    \newpage
\pagestyle{empty}
\counterwithin{equation}{section}
\renewcommand{\thefigure}{S\arabic{figure}}
\setcounter{figure}{0}

\section*{Supplementary Material} \label{sec:supplementaryMaterial}

\subsection*{Time-averaged Wall Shear Stress Gradient}

    To quantify the spatial variation of the \gls{tawss} in our study, the
    \gls{tawssg} field, \glsxtrshort{tawssg}, as defined by \citet{Geers2017},
    was computed:
    \begin{equation} \label{eq:timeAveragedWSSGradient}
        \gls{timeAvgWallShearStressGradient}
        =
        \left(
            \tgrad{\tavg{\gls{wallShearStressMag}}}
        \right)
        \dprod
        \gls{avgShearDirectionVector}\,,
    \end{equation}
    where $\tgrad{}$ is the tangential or surface gradient operator, given by:
    \begin{equation} \label{eq:tangentialGradientOperator}
        \tgrad{}
        \equiv
        \grad{}
        -
        \gls{sNormalVector}
        \left(
            \gls{sNormalVector}
            \dprod
            \grad{}
        \right)\,,
    \end{equation}
    with $\grad$ is the gradient operator in Cartesian coordinates and
    \gls{avgShearDirectionVector} is the unit vector pointing in the
    time-averaged \gls{wss} vector direction:
    \begin{equation} \label{eq:avgShearDirectionVector}
        \gls{avgShearDirectionVector}
        =
        \frac{
            \displaystyle
            \int_{0}^{\gls{cardiacPeriod}}
            {
                \gls{wallShearStressVector}
                \diff{\gls{time}}
            }
        }{
            \displaystyle
            \left \|
                \int_{0}^{ \gls{cardiacPeriod} }
                {
                    \gls{wallShearStressVector}
                    \left(
                        \gls{EulerCoord},
                        \gls{time}
                    \right)
                    \diff{\gls{time}}
                }
            \right \|
        }\,.
    \end{equation}

\begin{figure}[!htb]
        \caption{%
             (a) and (b) Correlations between the \gls{tawssg} and the
             surface-averaged Gaussian curvature and (c) distribution of the
             \gls{tawssg} for all aneurysms in our database, segregated by
             rupture-status groups.
        }

        \includegraphics[width=\textwidth]{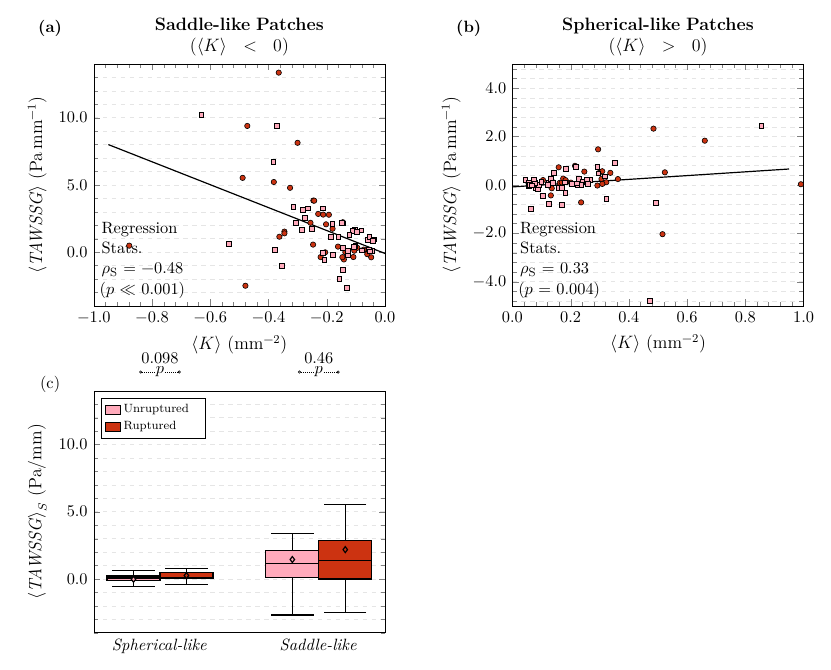}

        \label{fig:TAWSSGDistributions}
    \end{figure}

    In \cref{fig:TAWSSGDistributions}, we present the distribution of the
    \gls{tawssg} for all aneurysms in our database, segregated by
    rupture-status groups as also the correlations between the \gls{tawssg} and
    the surface-averaged Gaussian curvature.

\subsection*{Flow Temporal Evolution}

\begin{figure}[!p]
        \caption{%
            Evolution of the (a) \gls{lambda2VortexId} isosurface, (b)
            \gls{wss} magnitude, and (c) \gls{lambda2VortexId} field over the
            cardiac cycle for a representative \gls{ia} case (case C0002 from
            the Aneurisk database) at three instants of the cardiac cycle.
        }

        \includegraphics[width=\textwidth]{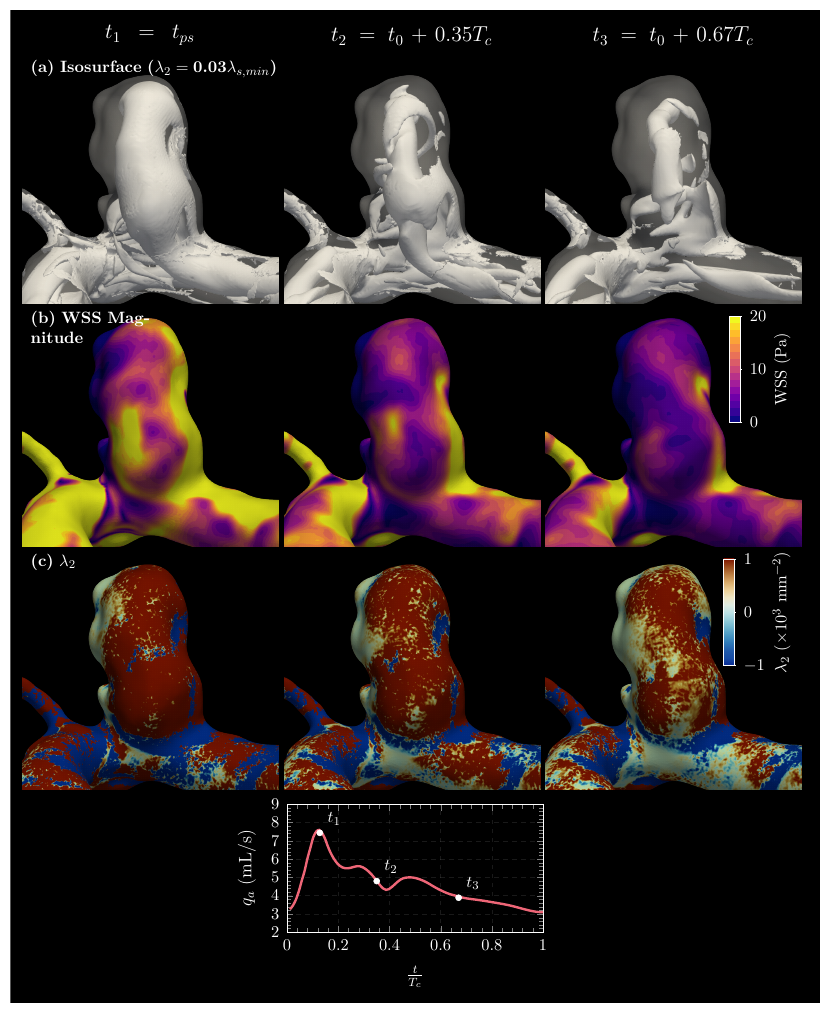}

        \label{fig:flowEvolution}
    \end{figure}

    To illustrate the temporal evolution of the flow structures and wall shear
    stress patterns in \glspl{ia}, we present in \cref{fig:flowEvolution} the
    evolution of the \gls{lambda2VortexId} isosurface, the \gls{wss} magnitude,
    and \gls{lambda2VortexId} field over the cardiac cycle for a representative
    \gls{ia} case (case C0002 from the Aneurisk database) at three instants of
    the cardiac cycle.

\end{document}